\documentclass[reprint,aip,pop,graphicx,numerical,amsfonts,amsmath,amssymb,twocolumn]{revtex4-1}

\usepackage{bm}
\usepackage[dvips]{graphicx}
\usepackage{lineno}
\usepackage{color}
\usepackage{soul}
\usepackage{booktabs}
\usepackage{multirow}
\sethlcolor{yellow}

\newcommand{\biota}{\iota
                     \hskip-.15ex{\hbox to 0pt{\hss {\leavevmode
                     \hbox{\raise -.60ex \hbox{{\tt \'{}}}}}}}
                     \hskip.37ex{\hbox to 0pt{\hss {\leavevmode
                     \hbox{\raise -.50ex \hbox{{\tt \'{}}}}}}}}

\begin{document}

\preprint{00}

\title{
Effects of collisions on conservation laws in gyrokinetic field theory
}



\author{H. Sugama}
\affiliation{
National Institute for Fusion Science, 
Toki 509-5292, Japan
}
\affiliation{
Department of Fusion Science, SOKENDAI (The Graduate University for Advanced Studies), 
Toki 509-5292, Japan 
}

\author{T.-H. Watanabe}
\affiliation{
Department of Physics,
Nagoya University,  
Nagoya 464-8602, Japan
}

\author{M. Nunami}
\affiliation{
National Institute for Fusion Science, 
Toki 509-5292, Japan
}
\affiliation{
Department of Fusion Science, SOKENDAI (The Graduate University for Advanced Studies), 
Toki 509-5292, Japan 
}


\date{\today}

\begin{abstract}
Effects of collisions on conservation laws for toroidal plasmas are investigated based on the gyrokinetic field theory. 
Associating the collisional system with a corresponding collisionless system at a given time such that the two systems have the same distribution functions and electromagnetic fields instantaneously, 
it is shown how the collisionless conservation laws derived from Noether's theorem are modified by the collision term. 
Effects of the external source term added into the gyrokinetic equation can be formulated similarly with the collisional effects. 
Particle, energy, and toroidal momentum balance equations including collisional and turbulent transport fluxes are systematically derived using a novel gyrokinetic collision operator, 
by which 
the collisional change rates of energy and canonical toroidal angular momentum 
per unit volume in the gyrocenter space 
can be given in the conservative forms. 
The ensemble-averaged transport equations of particles, energy, and toroidal momentum given in the present work are shown to include classical, neoclassical, and turbulent transport fluxes which agree with those derived from conventional recursive 
formulations. 
\end{abstract}

\pacs{52.25.Dg, 52.25.Fi, 52.25.Xz, 52.30.Gz, 52.35.Ra, 52.55.Dy, 52.55.F} 

\maketitle 



\section{INTRODUCTION}

Gyrokinetic theories and simulations are powerful means to 
investigate microinstabilities and turbulent transport processes 
in magnetically confined 
plasmas.~\cite{Krommes2012,Garbet2010,Idomura2006,dimits00} 
   Originally, gyrokinetic equations are derived by recursive 
techniques combined with the WKB or ballooning 
representation.~\cite{HM,rutherford,taylor,AL,CTB,Frieman} 
   On the other hand, modern derivations of the gyrokinetic equations 
are based on the Lagrangian and/or Hamiltonian formulations,~\cite{B&H} 
in which conservation laws for the phase-space volume and the magnetic 
moment are automatically ensured by 
Liouville's theorem and Noether's theorem, respectively.~\cite{goldstein} 
   Besides, conservation of the total energy and momentum 
is naturally obtained in the gyrokinetic field theory, 
where all governing equations for 
the distribution functions and the electromagnetic fields 
are derived from 
the Lagrangian which describes the whole system 
consisting of particles and 
fields.~\cite{Sugama2000,Brizard2000,Scott,Brizard2011,Squire}
   A subtle point regarding the Lagrangian/Hamiltonian gyrokinetic formulations 
is that they basically treat collisionless systems so that 
Noether's theorem and conservation laws do not hold directly for collisional systems.   
   In this paper, we examine how the collision and external source terms 
added into the gyrokinetic equations influence the conservation laws derived from Noether's theorem in the gyrokinetic field theory for collisionless systems. 

For a given collisional kinetic system, 
we can imagine a corresponding collisionless kinetic system 
such that the two systems have the same distribution functions 
and electromagnetic fields instantaneously. 
   As an example of two such systems, 
the Boltzmann-Poisson-Amp\`{e}re system 
and the Vlasov-Poisson-Amp\`{e}re system are considered in Sec.~II, 
where we express the variation of the action integral for the latter collisionless system 
in terms of 
the distribution functions and the electromagnetic fields for the 
former collisional system to show how the conservation laws 
derived from Noether's theorem in the collisionless system
are modified in the collisional system with  
external sources of particles, energy, and momentum. 
   There, we confirm the natural result that, 
when adding no external sources but only the collision term 
that conserves the energy and momentum,   
the energy and momentum conservation laws for  
the Boltzmann-Poisson-Amp\`{e}re system take the same forms 
as those for the Vlasov-Poisson-Amp\`{e}re system.
   The above-mentioned procedures are repeated in Sec.~III 
to treat the collisional and collisionless gyrokinetic systems. 
   In our previous work,~\cite{Sugama2014} 
using the gyrokinetic Vlasov-Poisson-Amp\`{e}re 
system of equations, 
conservation laws of particles, energy, and toroidal 
angular momentum  
are obtained for collisionless toroidal plasmas, 
in which the slow temporal variation of the background magnetic 
field is taken into account in order to enable self-consistent 
treatment of physical processes on transport time scales.  
  Based on these results, 
the particle, energy, and toroidal angular momentum 
balance equations for the collisional plasma are derived 
from the gyrokinetic Boltzmann-Poisson-Amp\`{e}re 
system of equations in Secs.~IV and V. 
  In Sec.~VI,  it is shown by 
taking the ensemble average of these balance equations 
that the particle, energy, and toroidal angular momentum 
transport fluxes are given by the sum of the 
conventional expressions of the classical, neoclassical, 
and turbulent transport fluxes 
to the lowest order in the normalized gyroradius parameter. 
   Conclusions are given in Sec.~VII and formulas for 
transformation from particle to gyrocenter coordinates are 
presented in Appendix~A. 

Regarding the collision operator for the gyrokinetic equation, 
several works have been done, which  
take account of finite-gyroradius effects to 
modify the Landau collision operator defined in
the particle 
coordinates.~\cite{Catto,Xu,Dimits,Brizard2004,Abel,Sugama2009,Madsen,Burby,Sugama_APSDPP}
   The relation of the collision operator in the gyrocenter coordinates 
to that in the particle coordinates is explained in Appendix~B. 
       The Landau operator for Coulomb collisions conserves  
particles' number, kinetic energy, and momentum locally at the particle position
although, in the gyrocenter position space, collisions induce transport fluxes 
of particles, energy, and momentum. 
    Besides, it is emphasized in this work that the collisional change rates 
of the gyrocenter Hamiltonian (which includes not only the kinetic energy 
but also the potential energy) and of the canonical momentum 
(instead of the kinetic momentum) per unit volume in the gyrocenter space 
should take the conservative (or divergence) forms in order to 
properly derive the energy and momentum conservation laws for 
the collisional gyrokinetic system. 
   The approximate collision operator which keeps these conservation 
properties of the gyrocenter energy and 
canonical toroidal angular momentum  
is shown in Appendix~C. 
%
[It is noted that another form of the gyrokinetic collision operator, 
which satisfies the energy and momentum conservation laws, has recently 
been presented by Burby {\it et al}.~\cite{Burby}]
  Appendix~D is given to describe how to derive the formula for 
the toroidal angular momentum transport flux due to the collision
term.

\section{Boltzmann-Poisson-Amp\`{e}re system}

In this section, conservation laws are investigated for 
the Boltzmann-Poisson-Amp\`{e}re system of 
equations which provide the basis 
of approximate description by the collisional electromagnetic gyrokinetic system of equations 
for strongly magnetized plasmas considered in Secs.~III--VI. 
   Time evolution of the distribution function $f_a({\bf x}, {\bf v}, t)$ 
for particle species $a$ is described by the Boltzmann kinetic equation,
\begin{eqnarray}
\label{Boltzmann}
& & 
\hspace*{-5mm}
\left[
\frac{\partial}{\partial t} +
{\bf v}\cdot \nabla
+ \frac{e_a}{m_a}
\left\{{\bf E}({\bf x},t) +
\frac{1}{c} {\bf v} \times
{\bf B} ({\bf x},t) \right\}
\cdot \frac{\partial}{\partial {\bf v}}
\right]
f_a({\bf x}, {\bf v}, t)
\nonumber \\ & & 
= {\cal K}_a ({\bf x}, {\bf v}, t) 
, 
\end{eqnarray}
where ${\cal K}_a ({\bf x}, {\bf v}, t) $ denotes the rate of change in the 
distribution function $f_a$ due to Coulomb collisions and 
it may also include other parts representing external particle, momentum, 
and/or energy sources if any. 
   The electromagnetic fields ${\bf E}({\bf x}, t)$ and ${\bf B}({\bf x}, t)$ are written as 
${\bf E} = - \nabla \phi - c^{-1} \partial {\bf A}/\partial t$ and 
${\bf B} = \nabla \times {\bf A}$, 
where the electrostatic potential $\phi$ and the vector potential ${\bf A}$ 
are determined by Poisson's equation, 
\begin{equation}
\label{poisson}
\nabla^2 \phi ({\bf x}, t)
= -4\pi \sum_a e_a
\int
f_a({\bf x}, {\bf v}, t)
d^3 {\bf v}
\equiv 
- 4 \pi \sigma
,
\end{equation}
and Amp\`{e}re's law, 
\begin{equation}
\label{ampere}
\nabla^2 {\bf A}({\bf x}, t)
=
 - \frac{4\pi}{c} {\bf j}_T
,
\end{equation}
respectively. 
Here, the Coulomb (or transverse) gauge condition $\nabla \cdot {\bf A} =0$ 
is used and the current density 
$
{\bf j} \equiv \sum_a e_a n_a {\bf u}_a 
\equiv \sum_a  e_a
\int
f_a({\bf x}, {\bf v}, t){\bf v}
d^3 {\bf v}
$ 
(or any vector field) is written as 
${\bf j} = {\bf j}_L + {\bf j}_T$, 
where 
$
{\bf j}_L \equiv - (4\pi)^{-1} \nabla \int d^3 {\bf x}' 
(\nabla'\cdot {\bf j})/|{\bf x}-{\bf x}'|
$
and 
$
{\bf j}_T \equiv  (4\pi)^{-1} \nabla \times (
 \nabla \times \int d^3 {\bf x}'\;  
{\bf j}/|{\bf x}-{\bf x}'|)
$
represent the longitudinal (or irrotational) and transverse (or solenoidal) 
parts, respectively.~\cite{jackson}
   Equations~(\ref{Boltzmann}), (\ref{poisson}), and 
(\ref{ampere}) are the governing equations for  
the Boltzmann-Poisson-Amp\`{e}re system. 

Suppose that $f_a$, $\phi$, and ${\bf A}$ which satisfy 
Eqs.~(\ref{Boltzmann})--(\ref{ampere})  are given. 
Then, for the electromagnetic fields 
${\bf E} = - \nabla \phi - c^{-1} \partial {\bf A}/\partial t$ and 
${\bf B} = \nabla \times {\bf A}$ given from $\phi$ and ${\bf A}$, 
we consider the distribution function $f_a^V$ which is 
the solution of the Vlasov equation, 
\begin{eqnarray}
\label{Vlasov}
& & 
\hspace*{-5mm}
\left[
\frac{\partial}{\partial t} +
{\bf v}\cdot \nabla
+ \frac{e_a}{m_a}
\left\{{\bf E}({\bf x},t) +
\frac{1}{c} {\bf v} \times
{\bf B} ({\bf x},t) \right\}
\cdot \frac{\partial}{\partial {\bf v}}
\right]
f_a^V({\bf x}, {\bf v}, t)
\nonumber \\ & & 
= 0 . 
\end{eqnarray}
   We also assume $f_a^V$ to coincide instantaneously with $f_a$ at a given time $t_0$ 
so that $f_a^V({\bf x}, {\bf v}, t_0) = f_a({\bf x}, {\bf v}, t_0)$. 
   Therefore, equations obtained from Eqs.~(\ref{poisson}) and (\ref{ampere}) 
with $f_a$ replaced by $f_a^V$ also hold at $t_0$. 
  In other words,  $f_a^V$, $\phi$, and ${\bf A}$ satisfy the 
Vlasov-Poisson-Amp\`{e}re 
system of equations at $t_0$. 
  Note that the Vlasov-Poisson-Amp\`{e}re system of equations can be derived from the variational 
principle using the action ${\cal I}$ defined by Eq.~(1) in Ref.~29 
where 
its variation $\delta {\cal I}$  
associated with infinitesimal transformations 
of independent and dependent variables
[see Eqs.~(10) and (15) in Ref.~29] are explicitly shown in order to apply 
Noether's theorem for obtaining conservation laws of energy and momentum. 
  Now, let us use $f_a^V$, $\phi$, and ${\bf A}$ to define the action integral ${\cal I}$ 
over 
a small time interval, $t_0 - h / 2 \leq t \leq t_0 + h / 2$, 
during which 
the Vlasov-Poisson-Amp\`{e}re system of equations are approximately satisfied 
by $f_a^V$, $\phi$, and ${\bf A}$ within the errors of order $h$. 
  Then, neglecting the errors of higher order in $h$, 
the variation $\delta {\cal I}$ can be written in the same form as  
in Eq.~(15) of Ref.~29, 
\begin{equation}
\label{delta_I}
\delta {\cal I}
=
- \int_{t_0 - h / 2}^{t_0 + h / 2}
 dt \int  d^3 {\bf x} \; 
\left[ 
\frac{\partial }{\partial t}\delta G_0^V ({\bf x}, t)
+ \nabla \cdot \delta {\bf G}^V ({\bf x}, t)
\right]
,
\end{equation}
where $\delta G_0^V$ and $\delta {\bf G}^V$ are written as  
\begin{eqnarray}
\label{GV}
\delta G_0^V ({\bf x}, t) 
& = & 
{\cal E}_c^V \; \delta t_E 
- {\bf P}_c^V \cdot \delta {\bf x}_E
, \nonumber \\ 
\delta {\bf G}^V ({\bf x}, t) 
& = & 
{\bf Q}_c^V \; \delta t_E 
- \mbox{\boldmath$\Pi$}_c^V \cdot \delta {\bf x}_E
+ {\bf S}_\phi \; \delta \phi 
- \mbox{\boldmath$\Sigma$}_A \cdot \delta {\bf A}
. \hspace*{7mm}
\end{eqnarray}
Here, $t_E$, $\delta {\bf x}_E$, $\delta \phi$, and $\delta {\bf A}$ represent 
variations of $t$, ${\bf x}$, $\phi$, and ${\bf A}$, respectively, 
while 
${\cal E}_c^V$, ${\bf P}_c^V$, ${\bf Q}_c^V$, $\mbox{\boldmath$\Pi$}_c^V$, 
${\bf S}_\phi$, and 
$\mbox{\boldmath$\Sigma$}_A$ are defined by 
\begin{eqnarray}
{\cal E}_c^V & = & 
\sum_a \int d^3 {\bf v} \; 
f_a^V ({\bf x}, {\bf v}, t)
\left( \frac{1}{2} m_a |{\bf v}|^2 + e_a \phi  \right)
\nonumber \\ & & \mbox{}
+ \frac{1}{8\pi} 
\left( - |\nabla \phi|^2 + |{\bf B}|^2 \right)
,
\nonumber \\
{\bf P}_c^V & = & 
\sum_a \int d^3 {\bf v} \; 
f_a^V ({\bf x}, {\bf v}, t)
\left( m_a {\bf v} + \frac{e_a}{c} {\bf A}  \right)
, 
\nonumber \\
{\bf Q}_c^V & = & 
\sum_a \int d^3 {\bf v} \; 
f_a^V ({\bf x}, {\bf v}, t)
\left( \frac{1}{2} m_a |{\bf v}|^2 + e_a \phi  \right)
{\bf v}
\nonumber \\ & & \mbox{}
+ \frac{1}{4\pi} 
\left(
\frac{\partial \phi}{\partial t} \nabla \phi
+ \frac{\lambda}{c}
\frac{\partial {\bf A}}{\partial t} 
- \frac{\partial {\bf A}}{\partial t}
\times {\bf B}
\right)
, 
\nonumber 
\end{eqnarray}
\begin{eqnarray}
\label{EPV}
\mbox{\boldmath$\Pi$}_c^V
& = & 
\sum_a \int d^3 {\bf v} \; 
f_a^V ({\bf x}, {\bf v}, t) {\bf v}
\left( m_a {\bf v} + \frac{e_a}{c} {\bf A}  \right)
\nonumber \\ & & \mbox{}
+ \frac{1}{8\pi} \left( |\nabla \phi|^2 - B^2 \right) 
{\bf I}
+ \frac{1}{4\pi} 
\left[ - (\nabla \phi) (\nabla \phi)
\right.
\nonumber \\ & & 
\left. \mbox{}
+ ( (\nabla {\bf A}) - (\nabla {\bf A})^T ) 
\cdot (\nabla {\bf A})^T
- \frac{\lambda}{c} 
(\nabla {\bf A})^T
\right]
,
\nonumber \\ 
{\bf S}_\phi 
& = & 
 - \frac{1}{4\pi} \nabla \phi
, \hspace*{2mm}
\mbox{and} 
\hspace*{2mm}
\mbox{\boldmath$\Sigma$}_A
=
\frac{1}{4\pi} 
\left( {\bf B}\times {\bf I} + 
\frac{\lambda}{c}
{\bf I}  \right)
, \hspace*{2mm}
\end{eqnarray}
respectively, where the superscript $T$ represents the transpose of the tensor
 and ${\bf I}$ denotes the unit tensor. 
    The field variable $\lambda$ which appears in Eq.~(\ref{EPV}) 
is introduced in Ref.~29 as the Lagrange undetermined 
multiplier to derive the 
Coulomb gauge condition $\nabla \cdot {\bf A} = 0$ 
and it is shown from Eq.~(8) in Ref.~29 that 
\begin{equation}
\label{lambda}
\nabla^2 \lambda 
= 4\pi \nabla \cdot {\bf j}
.
\end{equation}

   Suppose that the variations $t_E$, $\delta {\bf x}_E$, $\delta \phi$, 
and $\delta {\bf A}$ are such that $\delta {\cal I} = 0$ 
holds for an arbitrary ${\bf x}$-integral domain in Eq.~(\ref{delta_I}). 
Then, taking the small time interval limit $h \rightarrow +0$ in Eq.~(\ref{GV}), 
we find that the conservation law,  
\begin{equation}
\label{conservationV}
\left[
\frac{\partial }{\partial t}\delta G_0^V ({\bf x}, t)
\right]_{t = t_0}
+ \nabla \cdot \delta {\bf G}^V ({\bf x}, t_0)
=0
,
\end{equation}
should be satisfied. 
   This is the so-called  ``Noether's theorem.” 
   Recalling that 
$f_a^V ({\bf x}, {\bf v}, t_0) = f_a ({\bf x}, {\bf v}, t_0)$ and 
comparing Eqs.~(\ref{Boltzmann}) and (\ref{Vlasov}) at  $t = t_0$, 
we have 
\begin{equation}
\label{dfvdt}
\left[
\frac{\partial f_a^V ({\bf x}, {\bf v}, t)}{\partial t}
\right]_{t = t_0}
=
\left[
\frac{\partial f_a ({\bf x}, {\bf v}, t)}{\partial t}
\right]_{t = t_0}
- {\cal K}_a ({\bf x}, {\bf v}, t_0)
.
\end{equation}
    We now define ${\cal E}_c$, ${\bf P}_c$, ${\bf Q}_c$, and 
$\mbox{\boldmath$\Pi$}_c$ from 
${\cal E}_c^V$, ${\bf P}_c^V$, ${\bf Q}_c^V$, and 
$\mbox{\boldmath$\Pi$}_c^V$, respectively, by replacing $f_a^V$ 
with $f_a$ in Eq.~(\ref{EPV}). 
   Correspondingly, $\delta G_0$ and $\delta {\bf G}$ are defined from 
$\delta G_0^V$ and $\delta {\bf G}^V$ by replacing ${\cal E}_c^V$, ${\bf P}_c^V$, 
${\bf Q}_c^V$, and $\mbox{\boldmath$\Pi$}_c^V$ with 
${\cal E}_c$, ${\bf P}_c$, ${\bf Q}_c$, and 
$\mbox{\boldmath$\Pi$}_c$, respectively,  in Eq.~(\ref{EPV}). 
   These definitions immediately yield
$\delta {\bf G}^V ({\bf x}, t_0) =  \delta {\bf G} ({\bf x}, t_0)$  and
\begin{equation}
\label{dG0Vt}
\left[
\frac{\partial \delta G_0^V ({\bf x}, t)}{\partial t}
\right]_{t = t_0}
=
\left[
\frac{\partial \delta G_0 ({\bf x}, t))}{\partial t}
\right]_{t = t_0}
- \delta K_{G0} ({\bf x}, t_0)
,
\end{equation}
where  Eq.~(\ref{dfvdt}) is used and $\delta K_{G0}$ 
is defined by 
\begin{eqnarray}
\label{KG0}
\delta K_{G0} 
& = & 
 K_{{\cal E}c} \; \delta t_E 
- {\bf K}_{Pc} \cdot \delta {\bf x}_E
,
\nonumber \\ 
K_{{\cal E}c}
& = & 
\sum_a \int d^3 {\bf v} \; 
{\cal K}_a
\left( \frac{1}{2} m_a |{\bf v}|^2 + e_a \phi  \right)
,
\nonumber \\ 
{\bf K}_{Pc}
& = & 
\sum_a \int d^3 {\bf v} \; 
{\cal K}_a
\left( m_a {\bf v} + \frac{e_a}{c} {\bf A}  \right)
.
\end{eqnarray}
Substituting Eq.~(\ref{dG0Vt}) into Eq.~(\ref{conservationV}), 
we find that the conservation law is modified for the 
Boltzmann-Poisson-Amp\`{e}re system as 
\begin{equation}
\label{Kbalance}
\frac{\partial }{\partial t}\delta G_0 ({\bf x}, t)
+ \nabla \cdot \delta {\bf G} ({\bf x}, t)
=
\delta K_{G0}
,
\end{equation}
where $t_0$ is rewritten as $t$ because $t_0$ is an arbitrarily chosen time. 
  Equation~(\ref{Kbalance}) shows that  $\delta K_{G0}$ represents 
effects of ${\cal K}_a$ in Eq.~(\ref{Boltzmann}) on the conservation law.  
  If ${\cal K}_a$ is given by the Coulomb collision term only, 
$\delta K_{G0}$ defined by Eq.~(\ref{KG0}) vanishes because
the collision term conserves particles' number, momentum, and energy. 

Energy and momentum balance equations can be derived 
from Eq.~(\ref{Kbalance}) using symmetries of the system 
under infinitesimal time and space translations as shown later. 
   Before deriving them, we first consider the equation for the particle number 
density $n_a  \equiv \int f_a d^3 {\bf v}$ which is obtained 
by taking the velocity-space integral 
of Eq.~(\ref{Boltzmann}) as
\begin{equation}
\label{na_eq}
\frac{\partial n_a}{\partial t} 
+ \nabla \cdot ( n_a {\bf u}_a )
=
\int {\cal K}_a d^3 {\bf v}
. 
\end{equation}
   We hereafter assume that $\sum_a e_a \int {\cal K}_a d^3 {\bf v} = 0$, 
which means that 
the source terms ${\cal K}_a$ conserve electric charge even if 
$\int {\cal K}_a d^3 {\bf v} \neq 0$ for each species $a$. 
   This seems a reasonable assumption in consistency with 
Eqs.~(\ref{poisson}) and (\ref{ampere}) in which no external source 
terms are included. 
   Then, multiplying Eq.~(\ref{na_eq}) with the electric charge $e_a$ and 
performing the summation over species result in the charge conservation law, 
\begin{equation}
\label{charge_conservation}
\frac{\partial \sigma}{\partial t} 
+ \nabla \cdot {\bf j}
=
0
.
\end{equation}
   We also find from 
$\sum_a e_a \int {\cal K}_a d^3 {\bf v} = 0$ that, 
in Eq.~(\ref{KG0}), 
the terms including $\phi$ and ${\bf A}$ vanish 
and make no contribution 
to $K_{{\cal E}c}$ and ${\bf K}_{Pc}$. 
    As seen from Eqs.~(\ref{poisson}), (\ref{lambda}), 
and (\ref{charge_conservation}), 
we can put $\lambda = \partial \phi / \partial t$ which is also used in Ref.~29 
to derive energy and momentum conservation laws for the Vlasov-Poisson-Amp\`{e}re system. 

   We now note that the action integral is invariant, namely,  $\delta {\cal I} = 0$ 
under the infinitesimal translations in space and time represented by 
$\delta t_E = \epsilon_0$, 
$\delta {\bf x}_a = \delta {\bf x}_E = 
\mbox{\boldmath$\epsilon$}$, 
$\delta {\bf v}_a = 0$, 
$\delta \phi = 0$, and 
$\delta {\bf A} = 0$, 
where $\epsilon_0$ and $\mbox{\boldmath$\epsilon$}$ are constant in time and space. 
   These invariance properties hold because 
the integrands in the action integral ${\cal I}$ depend on $({\bf x}, t)$ only through  variational variables [see Eq.~(1) in Ref.~29].  
   Using the time translational symmetry, Eq.~(\ref{Kbalance}) reduces to 
the canonical energy balance equation, 
\begin{equation}
\label{cons_e1}
\frac{\partial {\cal E}_c}{\partial t} 
+ \nabla \cdot {\bf Q}_c
=
K_{{\cal E}c}
, 
\end{equation}
where the canonical energy density and flux $({\cal E}_c, {\bf Q}_c)$ are 
given by replacing $f_a^V$ with $f_a$ in the definitions of $({\cal E}_c^V, {\bf Q}_c^V)$ in Eq.~(\ref{EPV}). 
    In the same way as in Ref.~29, 
we use the kinetic energy density and flux,  $({\cal E}_p, {\bf Q}_p)$, 
defined by 
\begin{eqnarray}
\label{EpQp}
{\cal E}_p 
& = &
\sum_a \int d^3 {\bf v} \; 
f_a ({\bf x}, {\bf v}, t)
\frac{1}{2} m_a |{\bf v}|^2 
, \nonumber \\
{\bf Q}_p 
& = &
\sum_a \int d^3 {\bf v} \; 
f_a ({\bf x}, {\bf v}, t)
\frac{1}{2} m_a |{\bf v}|^2  {\bf v}
, 
\end{eqnarray}
    to modify Eq.~(\ref{cons_e1}) into more familiar forms. 
    Then, 
the energy balance equation is finally written as 
\begin{eqnarray}
\label{cons_e2}
& & 
\frac{\partial }{\partial t}
\left( {\cal E}_p + 
\frac{|{\bf E}_L|^2 + |{\bf B}|^2 }{8\pi} \right)
\nonumber \\ 
& & \mbox{}
+ \nabla \cdot \left( {\bf Q}_p 
+ \frac{c}{4\pi} {\bf E}\times {\bf B}
- \frac{1}{4\pi} \frac{\partial \phi}{\partial t}
{\bf E}_T \right)
\nonumber \\ 
& = &
\frac{\partial }{\partial t}
\left( {\cal E}_p + 
\frac{|{\bf E}_L|^2 + 2 {\bf E}_L \cdot {\bf E}_T 
+ |{\bf B}|^2 }{8\pi}\right)
\nonumber \\ & & \mbox{}
+ \nabla \cdot \left( {\bf Q}_p 
+ \frac{c}{4\pi} {\bf E}\times {\bf B}
+ \frac{1}{4\pi} \phi \frac{\partial {\bf E}_T}{\partial t}
\right)
\nonumber \\ 
& = &
K_{{\cal E}c}
\equiv
\sum_a \int d^3 {\bf v} \; 
{\cal K}_a
 \frac{1}{2} m_a |{\bf v}|^2 
,
\end{eqnarray}
where ${\bf E}_L \equiv - \nabla \phi$ and 
${\bf E}_T \equiv - c^{-1} \partial {\bf A}/\partial  t$ are 
the longitudinal and transverse parts of 
the electric field, respectively. 

Next, from the space translational symmetry and Eq.~(\ref{Kbalance}), 
we obtain the canonical momentum balance equation, 
\begin{equation}
\label{cons_m1}
\frac{\partial {\bf P}_c}{\partial t} 
+ \nabla \cdot \mbox{\boldmath$\Pi$}_c
=
{\bf K}_{Pc}
,
\end{equation}
where the canonical momentum density and 
tensor $({\bf P}_c, \mbox{\boldmath$\Pi$}_c)$ are 
given by replacing $f_a^V$ with $f_a$ in the definitions 
of $({\bf P}_c^V, \mbox{\boldmath$\Pi$}_c^V)$ in Eq.~(\ref{EPV}). 
   Furthermore,  in the same way as in Ref.~29,  
the invariance of $I$ under the infinitesimal rotation is shown to 
give the equation for the angular momentum, which is used to 
modify Eq.~(\ref{cons_m1}) into the momentum balance equation, 
\begin{equation}
\label{cons_m4}
\frac{\partial }{\partial t}
({\bf P}_p + {\bf P}_f)
+ \nabla \cdot ( \mbox{\boldmath$\Pi$}_p + \mbox{\boldmath$\Pi$}_f )
 = 
{\bf K}_{Pc}
\equiv 
\sum_a \int d^3 {\bf v} \; 
{\cal K}_a
 m_a {\bf v} 
.
\end{equation}
   Here, the particle parts $({\bf P}_p, \mbox{\boldmath$\Pi$}_p)$ of the 
momentum density and the pressure tensor are defined by 
\begin{eqnarray}
{\bf P}_p 
& = & 
\sum_a \int d^3 {\bf v} \; f_a ({\bf x}, {\bf v}, t)
m_a {\bf v}
,
\nonumber \\ 
\mbox{\boldmath$\Pi$}_p
& = & 
\sum_a \int d^3 {\bf v} \; f_a ({\bf x}, {\bf v}, t)
m_a {\bf v} {\bf v}
,
\end{eqnarray}
and the field parts $({\bf P}_f, \mbox{\boldmath$\Pi$}_f)$ 
are given by
\begin{eqnarray}
{\bf P}_f 
& = & 
\frac{{\bf E}_L \times {\bf B}}{4\pi c}
,
\nonumber \\ 
\mbox{\boldmath$\Pi$}_f
& = & 
\frac{1}{8\pi}
(|{\bf E}_L|^2 + 2 {\bf E}_L \cdot {\bf E}_T
+ |{\bf B}|^2 ){\bf I}
\nonumber \\ & &  \mbox{}
- \frac{1}{4\pi}
( {\bf E}_L {\bf E}_L + {\bf E}_L {\bf E}_T
+ {\bf E}_T {\bf E}_L + {\bf B} {\bf B} )
.
\end{eqnarray}

Equations~(\ref{cons_e2}) and (\ref{cons_m4}) take physically familiar forms of energy and momentum balance equations including external source terms. 
   As mentioned earlier, if ${\cal K}_a$ is given by the Coulomb collision term only, 
$K_{{\cal E}c}$ and ${\bf K}_{Pc}$ vanish so 
that the energy and momentum balance equations for the 
Boltzmann-Poisson-Amp\`{e}re system take the same forms as those for 
the Vlasov-Poisson-Amp\`{e}re system.~\cite{Sugama2013} 

\section{Gyrokinetic Boltzmann-Poisson-Amp\`{e}re  system}

   Let us start from the gyrokinetic Boltzmann equation 
written as 
\begin{eqnarray}
\label{gkb}
& & 
\left(
\frac{\partial}{\partial t} 
+   
\frac{d {\bf Z}_a}{dt} 
\cdot
 \frac{\partial}{\partial {\bf Z}} 
\right)
 F_a ({\bf Z}, t) 
\nonumber \\
& = & 
\sum_b C_{ab}^g [F_a, F_b] ({\bf Z}, t) + {\cal S}_a ({\bf Z}, t)
, 
\end{eqnarray}
   where $F_a({\bf Z}, t)$ is the gyrocenter distribution function 
for species $a$,  $C_{ab}^g [F_a, F_b] ({\bf Z}, t)$ represents 
the rate of change in $F_a({\bf Z}, t)$ due to Coulomb 
collisions between particle species $a$ and $b$, and 
 ${\cal S}_a ({\bf Z}, t)$ denotes other parts including    
external particle, momentum, and/or energy sources if any.
   The gyrocenter coordinates are written as  
${\bf Z}_a = ({\bf X}_a, U_a, \mu_a, \xi_a)$, where 
${\bf X}_a$, $U_a$, $\mu_a$, and $\xi_a$ represent
the gyrocenter position, parallel velocity, magnetic moment, and 
gyrophase angle, respectively. 
   Appendix~A shows the relation of the gyrocenter coordinates 
to the particle coordinates in detail. 
   The perturbation expansion parameter in the gyrokinetic theory 
is denoted by $\delta$ which represents the ratio of the gyroradius $\rho$ 
to the macroscopic scale length $L$ of the background field. 
   It is shown in Appendix~B how the collision operator 
$C_{ab}^g [F_a, F_b]$ 
for the gyrocenter distribution functions $F_a$ and $F_b$ 
is given  
from the collision operator 
$C_{ab}^p [f_a, f_b]$
for the particle distribution functions $f_a$ and $f_b$.  

   The deviation of each distribution function from the 
local Maxwellian is regarded as of ${\cal O}(\delta)$, and accordingly 
the collision term $C_{ab}^g$ is considered to be of ${\cal O}(\delta)$. 
  We assume that the source term  ${\cal S}_a$ is of ${\cal O}(\delta^2)$ 
so that its effect appears only in the transport time scale.  
  We also assume that 
$\sum_a e_a \int dU \int d\mu \int d\xi \; D_a {\cal S}_a ({\bf Z}, t) = 0$ 
in order to prevent the source term from affecting the charge conservation laws 
[see Eq.~(\ref{gccc})]. 
   Here, $D_a$ denotes 
the Jacobian for the gyrocenter coordinates, 
$
D_a \equiv 
\det [ \partial ({\bf x}_a, {\bf v}_a)/  
\partial ({\bf X}_a, U_a, \xi_a,  \mu_a ) ]
$, 
where  $({\bf x}_a, {\bf v}_a)$ represent 
the particle coordinates consisting of the particle's position and velocity vectors. 

    We treat toroidal systems, for which the equilibrium magnetic field 
is given in the axisymmetric form as 
\begin{equation}
\label{B0}
{\bf B}_0 = \nabla \times {\bf A}_0 = 
I \nabla \zeta 
+ \nabla \zeta \times \nabla \chi
,
\end{equation}
where $I$ and $\chi$ are constant on toroidal flux surfaces labeled by an 
arbitrary radial coordinate $s$ and $\zeta$ is the toroidal angle.   
    We note that $I$ and $\chi$ represent the covariant toroidal component of 
the equilibrium field ${\bf B}_0$ and the poloidal magnetic flux divided 
by $2\pi$, respectively. 
   The equilibrium field ${\bf B}_0$ is allowed to be dependent on time. 
   Then, following Ref.~18,
the gyrocenter motion equations are written as  
\begin{equation}
\label{dZdt}
\frac{d {\bf Z}_a}{dt} = 
\left\{ {\bf Z}_a, H_a \right\}
+ \left\{ {\bf Z}_a, {\bf X}_a  
\right\}
\cdot \frac{e_a}{c}
\frac{\partial {\bf A}_a^*}{\partial t}
,
\end{equation}
where the gyrocenter Hamiltonian, which is independent of $\xi_a$, 
is defined by 
\begin{equation}
\label{Ha}
H_a 
=  \frac{1}{2} m_a U_a^2 + \mu_a B_0  
+ e_a \Psi_a 
,  
\end{equation}
%
   and ${\bf A}_a^{*}$ is given by 
$
{\bf A}_a^{*}  
= 
{\bf A}_0 ({\bf X}_a, t) 
+ (m_a c / e_a)
 U_a {\bf b}
 ({\bf X}_a, t)
$.
    Using the nonvanishing components of the Poisson brackets for pairs of the 
gyrocenter coordinates given by 
\begin{eqnarray}
\label{PB}
& & \{ {\bf X}_a,  {\bf X}_a \} 
= \frac{c}{e_a B_{a\parallel}^*} 
{\bf b} \times {\bf I}
,
\hspace*{5mm}
\{ {\bf X}_a,  U_a \} 
=
\frac{{\bf  B}_a^*}{m_a B_{a\parallel}^*} 
,
\nonumber \\ & & 
\{ \xi_a, \mu_a \} 
= \frac{e_a}{m_a  c}
, 
\end{eqnarray}
  the gyrocenter motion equations in Eq.~(\ref{dZdt}) 
are rewritten as  
\begin{eqnarray}
\label{dXdt}
\frac{d {\bf X}_{a}}{dt}
&  = & 
\frac{1}{B^*_{a\parallel}}
\left[
\left(U_{a}+  \frac{e_a}{m_a}
\frac{\partial \Psi_a}{
\partial U_a } \right) {\bf B}^*_a
\right. \nonumber \\ & & 
\left. \mbox{}
+  c{\bf b}\times
\left(\frac{\mu_{a}}{e_a} \nabla B_0
+ \nabla \Psi_a 
+ \frac{1}{c} \frac{\partial {\bf A}_a^*}{\partial t}
\right)
\right]
,
\end{eqnarray}
\begin{equation}
\label{dUdt}
\frac{d U_{a}}{dt}
 = 
- \frac{{\bf B}^*_a}{m_a B^*_{a\parallel}}
\cdot
\left[\mu_{a} \nabla B_0
+  e_a 
\left( \nabla \Psi_a 
+ \frac{1}{c} \frac{\partial {\bf A}_a^*}{\partial t} 
\right) \right]
,
\end{equation}
\begin{equation}
\label{dmu}
\frac{d \mu_{a}}{dt}
 = 0
,
\end{equation}
and
\begin{equation}
\label{dxi}
\frac{d \xi_{a}}{dt}
 = 
\Omega_a
+
\frac{e_a^2}{m_a c}
\frac{\partial\Psi_a}{
\partial \mu_{a}}
. 
\end{equation}
   Here, $\Omega_a \equiv e_a B_0/(m_a c)$, ${\bf b} = {\bf B}_0 / B_0$, 
$B_{a\parallel}^* = {\bf B}_a^* \cdot {\bf b}$, 
and 
${\bf B}_a^* = \nabla \times  {\bf A}_a^*$.  
   The field variable $\Psi_a$ is defined by 
\begin{eqnarray}
\label{Psi}
\Psi_a 
& = & 
\left\langle \psi_a({\bf Z}_a,t) \right\rangle_{\xi_a}
+ \frac{e_a}{2m_a c^2}  \left\langle 
|{\bf A}_1 
({\bf X}_a + \mbox{\boldmath $\rho$}_a,t) 
|^2 \right\rangle_{\xi_a}
\nonumber \\ & & \mbox{}
- \frac{e_a}{2 B_0}
\frac{\partial}{\partial \mu} \langle 
[
\widetilde{\psi}_a({\bf Z}_a,t) 
]^2
\rangle_{\xi_a}
,  
\end{eqnarray}
where the field variable $\psi_a$ is defined  in terms of  
 the electrostatic potential $\phi$ and 
the perturbation part of the vector potential ${\bf A}_1$ as 
\begin{equation}
\label{psi=}
\psi_a({\bf Z}_a,t) 
= \phi ({\bf X}_a + \mbox{\boldmath $\rho$}_a,t) 
- \frac{1}{c}{\bf v}_{a0}({\bf Z}_a, t)\cdot {\bf A}_1
({\bf X}_a + \mbox{\boldmath $\rho$}_a,t)
.
\end{equation}
   The gyroradius vector is given by 
$\mbox{\boldmath $\rho$}_a  
= {\bf b}({\bf X}_a, t) \times {\bf v}_{a0} ({\bf Z}_a, t)
/ \Omega_a({\bf X}_a, t)$ and 
the zeroth-order particle velocity 
${\bf v}_{a0}$ is written in terms of the gyrocenter coordinates as 
$
{\bf v}_{a0}({\bf Z}_a, t) 
=
U_a {\bf b} ({\bf X}_a, t) - 
[2  \mu_a B_0({\bf X}_a)/m_a]^{1/2}
[ \sin \xi_a  {\bf e}_1({\bf X}_a, t)
+ \cos \xi_a  {\bf e}_2({\bf X}_a, t)]
$, 
where the unit vectors $({\bf e}_1, {\bf e}_2, {\bf b})$ form a right-handed orthogonal system. 
   The gyrophase-average and gyrophase-dependent parts of 
an arbitrary periodic function $Q(\xi_a)$ of the gyrophase $\xi_a$ 
are written as  
\begin{equation}
\langle Q \rangle_{\xi_a} 
\equiv \oint \frac{d\xi_a}{2\pi} Q (\xi_a)
\hspace*{3mm}
\mbox{and}
\hspace*{3mm}
\widetilde{Q}
\equiv Q - \langle Q \rangle_{\xi_a} 
,
\end{equation}
respectively. 
   In the gyrocenter motion equations, 
effects of the time-dependent 
background magnetic field and those of the fluctuating 
electromagnetic fields appear through 
$\partial {\bf A}_a^* / \partial t$ and  $\Psi_a$, 
respectively.
   It should be noted that  
$d {\bf Z}_a /d t$ on the left-hand side of Eq.~(\ref{gkb}) 
is regarded as a function of $({\bf Z}, t)$ which 
is given by the right-hand side of Eq.~(\ref{dZdt}). 

   We find from Eqs.~(\ref{Deltaz})--(\ref{Z=Tz}) in Appendix~A and 
Eq.~(\ref{scalar_C}) in Appendix~B that 
the gyrophase-dependent part of the right-hand side 
of Eq.~(\ref{gkb}) appears from $C_{ab}^g$ and it is of ${\cal O}(\delta)$. 
   Using $\Omega_a = {\cal O}(\delta^{-1})$, 
the gyrophase-dependent part of the left-hand side of Eq.~(\ref{gkb}) 
is written as $\Omega_a \partial \widetilde{F}_a/\partial \xi$ 
to the lowest order in $\delta$. 
   Then, it is concluded that $\widetilde{F}_a = {\cal O} (\delta^2)$. 
   Taking the gyrophase average of Eq.~(\ref{gkb}), we obtain 
\begin{eqnarray}
\label{gkbav}
& & 
\left(
\frac{\partial}{\partial t} 
+   
\frac{d {\bf Z}_a}{dt} 
\cdot
 \frac{\partial}{\partial {\bf Z}} 
\right)
F_a ({\bf Z}, t) 
\nonumber \\
& = & 
\sum_a
\langle C_{ab}^g  [F_a, F_b] ({\bf Z}, t) \rangle_\xi
+ {\cal S}_a ({\bf Z}, t) 
, 
\end{eqnarray}
    where $F_a ({\bf Z}, t)$ and ${\cal S}_a ({\bf Z}, t)$ 
are both regarded as independent of the gyrophase $\xi$ 
and $\langle \cdots \rangle_\xi$ are omitted from them for simplicity.
       It is seen from Eq.~(\ref{scalar_C}) that 
effects of $\widetilde{F}_a = {\cal O} (\delta^2)$ on 
$\langle C_{ab}^g  [F_a, F_b] \rangle_\xi$ in the right-hand side 
of Eq.~(\ref{gkbav})
are estimated as of ${\cal O} (\delta^3)$. 
    Here and hereafter, we neglect 
$\widetilde{F}_a = {\cal O} (\delta^2)$ in both sides of 
the gyrokinetic Boltzmann equation 
given by Eq.~(\ref{gkbav}). 
   Even so, its moment equations can correctly include
the collisional transport fluxes of particles, energy, and 
toroidal momentum up to the leading order, that is  
${\cal O}(\delta^2)$, as confirmed later. 
   In Appendix~C, Eq.~(\ref{DCg2}) combined with  
Eqs.~(\ref{DvDm}), (\ref{aa1}), (\ref{aa2}), and (\ref{aa3})  
presents the approximate gyrokinetic collision operator, 
which has favorable conservation properties 
and correctly 
describes collisional transport of energy and toroidal 
angular momentum. 

     The gyrokinetic Poisson equation and 
the gyrokinetic Amp\`{e}re's law are written as~\cite{Sugama2014}
\begin{eqnarray}
\label{gkp}
\nabla^2 \phi ({\bf x}, t)
 & =  & -4\pi \sum_a e_a 
\int d^6 {\bf Z} \; D_a ({\bf Z}, t) 
\delta^3 ( {\bf X} + \mbox{\boldmath $\rho$}_{a}
-{\bf x} )
\nonumber \\ & & \mbox{}
\times
\left[
F_a ({\bf Z}, t) 
+ \frac{e_a \widetilde{\psi}_a}{B_0}
\frac{\partial F_a}{\partial \mu}
\right]
, 
\end{eqnarray}
and 
\begin{equation}
\label{ampere1b}
\nabla^2 
( {\bf A}_0 + {\bf A}_1 )  =  
- \frac{4\pi}{c} 
({\bf j}_G)_T
,
\end{equation}
respectively,  where $({\bf j}_G)_T$ is the transverse part of 
the gyrokinetic current density ${\bf j}_G$ defined by 
\begin{eqnarray}
\label{jg}
{\bf j}_G
& \equiv  & 
\sum_a  e_a
\int d^6 {\bf Z} D_a ({\bf Z})
\delta^3[ {\bf X} +
\mbox{\boldmath $\rho$}_{a}({\bf Z})
 -{\bf x} ]
\nonumber \\ & & \mbox{} \times
\left(
F_a ({\bf Z}, t)
\left[
{\bf v}_{a0} ({\bf Z})
- \frac{e_a}{m_a c} {\bf A}_1
({\bf X} + 
\mbox{\boldmath $\rho$}_{a}({\bf Z}) , t)
\right]
\right. 
\nonumber \\ & & 
\left. \mbox{}
+ 
\frac{e_a \widetilde{\psi}_a}{B_0}
\frac{\partial F_a}{\partial \mu}
{\bf v}_{a0} ({\bf Z})
\right)
.
\end{eqnarray} 
    It should also be noted that 
the Coulomb gauge conditions $\nabla \cdot {\bf A}_0 = 0$ and 
$\nabla \cdot {\bf A}_1 = 0$ for the equilibrium and perturbation parts of 
the vector potential are used here. 
     The equilibrium vector potential ${\bf A}_0$ is given by 
$
{\bf A}_0 
=
-\chi \nabla \zeta + \nabla \zeta \times \nabla \eta
$,
   where 
$
\eta = \eta (R, Z)
$
   is the solution of 
$
\Delta_* \eta
\equiv R^2 \nabla \cdot (R^{-2} \nabla \eta)
= I
$.
   In Ref.~18, additional governing equations are derived 
in order to self-consistently determine $I$ and $\chi$ for 
the time-dependent axisymmetric background 
field ${\bf B}_0 = \nabla \times {\bf A}_0 
= I \nabla \zeta + \nabla \zeta \times \nabla \chi$.
   They are given by 
\begin{equation}
\label{I=}
 I =
\oint \frac{d\theta}{2\pi}
\left[
 \frac{4\pi}{c} \overline{M_\zeta} + \overline{(B^{\rm (gc)})_\zeta}
- \overline{B_{1\zeta}} \right]
, 
\end{equation}
   and
\begin{eqnarray}
\label{Dchi}
 \Delta_* \chi 
& = &
\overline{
\left(
\frac{4\pi}{c} 
\left[ ({\bf j}^{\rm (gc)})_T +
\nabla \times {\bf M} \right]  
 - \nabla \times {\bf B}_1 
\right)
\cdot R^2 \nabla \zeta}
\nonumber \\ & & \mbox{}
+ \frac{\partial I}{\partial \chi} 
\overline{\Lambda_\zeta}
,
\end{eqnarray}
   where the toroidal-angle average is represented by 
$\overline{\cdots} \equiv (2\pi)^{-1}\oint \cdots d\zeta$, 
the poloidal angle is denoted by $\theta$, and 
$
\overline{\Lambda_\zeta}
=  I - (4\pi/c) \overline{M_\zeta} - \overline{(B^{\rm (gc)})_\zeta} + \overline{B_{1\zeta}}
$.
   Here, the covariant toroidal component of an arbitrary vector ${\bf V}$ is written as $V_\zeta$. 
The turbulent part of the magnetic field is given by ${\bf B}_1 = \nabla \times {\bf A}_1$ and 
${\bf B}^{\rm (gc)}$ is defined by
$
\nabla \times {\bf B}^{\rm (gc)}
=
(4\pi/c)
({\bf j}^{\rm (gc)})_T
$.
   The magnetization ${\bf M}$ can be obtained from the 
turbulent fields and the distribution functions for 
all species using Eqs.~(41)--(43) in Ref.~18. 
   Thus, Eqs.~(\ref{gkbav}), (\ref{gkp}), (\ref{ampere1b}), (\ref{I=}), 
and (\ref{Dchi}) 
constitute the closed system of governing equations which determine $F_a$, 
$\phi$, ${\bf A}_1$, $I$, and $\chi$. 

For the gyrocenter coordinates which have Poisson brackets given by 
Eq.~(\ref{PB}), 
the Jacobian is given by 
$
D_a =
B_{a\parallel}^*/m_a
$.
 It is important to note that the Jacobian $D_a$
 satisfies 
the gyrocenter phase-space conservation law,   
\begin{equation}
\label{delD}
\frac{\partial D_a}{\partial t} + 
 \frac{\partial}{\partial {\bf Z}} \cdot 
\left( D_a 
\frac{d {\bf Z}_a}{dt} \right) 
=0
.
\end{equation}
    Then, using Eq.~(\ref{delD}), 
the gyrokinetic Boltzmann equation in Eq.~(\ref{gkbav}) 
can be rewritten as 
\begin{equation}
\label{gkb1}
\frac{\partial}{\partial t} 
\left( D_a F_a \right) 
+ \frac{\partial}{\partial {\bf Z}} 
\cdot
\left( D_a F_a 
\frac{d {\bf Z}_a}{dt}  \right)  
= 
D_a {\cal K}_a 
, 
\end{equation}
where ${\cal K}_a$ is the gyrophase-independent function 
given by the right-hand side of Eq.~(\ref{gkbav}),  
\begin{equation}
\label{gkK}
{\cal K}_a ({\bf Z}, t) 
=
\sum_a
\langle C_{ab}^g  [F_a, F_b] ({\bf Z}, t) \rangle_\xi
+ {\cal S}_a ({\bf Z}, t) 
.
\end{equation}

   We hereafter derive conservation laws for 
the gyrokinetic Boltzmann-Poisson-Amp\`{e}re  system of equations 
following the procedures similar to those shown in Sec.~II.  
    For that purpose, suppose that $F_a$, $\phi$, ${\bf A}_1$, $I$, and $\chi$ satisfy 
Eqs.~(\ref{gkbav}), (\ref{gkp}), (\ref{ampere1b}), (\ref{I=}), 
and (\ref{Dchi}). 
    Then, we consider the gyrocenter distribution function $F_a^V$ 
which obeys the gyrokinetic Vlasov equation,
\begin{equation}
\label{gkv}
\left(
\frac{\partial}{\partial t} 
+ \frac{d {\bf Z}_a}{dt}  \cdot
\frac{\partial}{\partial {\bf Z}} 
 \right)
F_a^V
=0
,
\end{equation}              
where $d {\bf Z}_a/dt$ is evaluated by using  
the above-mentioned fields $(\phi, {\bf A}_1, I, \chi)$
obtained from the solution of the 
gyrokinetic Boltzmann-Poisson-Amp\`{e}re  system of equations. 
    Here, it should be noted that, if 
the distribution functions $F_a$ and $F_a^V$, which are given as the solutions of 
Eqs.~(\ref{gkbav}) and (\ref{gkv}), respectively, are initially gyrophase-independent, 
they are gyrophase-independent at any time. 
    Besides, $F_a^V$ is assumed to coincide instantaneously 
with $F_a$ at a given time $t_0$. 
    Therefore,  
Eqs.~(\ref{gkp}), (\ref{ampere1b}), (\ref{I=}),  
and (\ref{Dchi}) are all satisfied at that moment even if  
$F_a$ is replaced with $F_a^V$ in these equations. 
   Thus, the gyrokinetic Vlasov-Poisson-Amp\`{e}re system of equations 
are instantaneously satisfied by $(F_a^V, \phi, {\bf A}_1, I, \chi)$ 
at $t=t_0$. 
   In Ref.~18, the action integral ${\cal I}$ 
is defined to derive all the governing equations for the 
gyrokinetic Vlasov-Poisson-Amp\`{e}re system based on the 
variational principle, and its variation $\delta {\cal I}$ associated with 
the infinitesimal variable transformations are given  
to obtain conservation laws from Noether's theorem. 
   Here, the action integral ${\cal I}$ can be expressed in terms of 
$(F_a^V, \phi, {\bf A}_1, I, \chi)$  over 
a small time interval, $t_0 - h / 2 \leq t \leq t_0 + h / 2$, 
during which the gyrokinetic 
Vlasov-Poisson-Amp\`{e}re system of equations are approximately satisfied 
by them within the errors of order $h$. 
  Then, neglecting the errors of higher order in $h$, 
we can write the variation $\delta {\cal I}$ in the same form as in Eq.~(77) 
of Ref.~18, 
\begin{equation}
\label{dIG}
\delta {\cal I}  =
- \int_{t_0 - h/2}^{t_0 + h/2} dt \int  d^3 {\bf X} \; 
\left[ 
\frac{\partial }{\partial t}\delta G_0^V ({\bf X}, t)
+ \nabla \cdot \delta {\bf G}^V ({\bf X}, t)
\right]
, 
\end{equation}
   with the functions $\delta G_0^V$ and $\delta {\bf G}^V$ defined by  
\begin{eqnarray}
\label{G}
\delta G_0^V ({\bf X}, t) 
& = & 
{\cal E}_c^V \; \delta t_E 
- {\bf P}_c^V \cdot \delta {\bf x}_E
, \nonumber \\
\delta {\bf G}^V ({\bf X}, t) 
& = & 
{\bf Q}_c^V \; \delta t_E 
- \mbox{\boldmath$\Pi$}_c^V \cdot \delta {\bf x}_E
+ {\bf S}_{\phi} \; \delta \phi 
- \mbox{\boldmath$\Sigma$}_{A1} \cdot \delta {\bf A}_1
\nonumber \\ & & \mbox{}
- \mbox{\boldmath$\Sigma$}_{A0}^V \cdot \delta {\bf A}_0
+ {\bf S}_\chi \delta \chi 
+ \delta {\bf T}^V
,
\end{eqnarray}
where  ${\cal E}_c^V$ and ${\bf P}_c^V $ are defined by 
\begin{eqnarray}
\label{EP}
{\cal E}_c^V & = & 
\sum_a \int dU \int d\mu \int d\xi \; 
D_a F_a^V H_a 
\nonumber \\ & & \mbox{}
+ \frac{1}{8\pi} 
\left( - |\nabla \phi|^2 
+ |{\bf B}_0 + {\bf B}_1 |^2 \right)
,
\nonumber \\
{\bf P}_c^V & = & 
\sum_a \int dU \int d\mu \int d\xi \; 
D_a F_a^V 
\left( m_a U {\bf b} + \frac{e_a}{c} {\bf A}_0  \right)
, 
\hspace*{7mm}
\end{eqnarray}
   and definitions of other variables 
${\bf Q}_c^V$, 
$\mbox{\boldmath$\Pi$}_c^V$,
${\bf S}_\phi$, 
$\mbox{\boldmath$\Sigma$}_{A1}$,
$\mbox{\boldmath$\Sigma$}_{A0}^V$
${\bf S}_\chi$, and 
$\delta {\bf T}^V$ 
are shown in Eq.~(79) of Ref.~18. 
   The superscript $V$ in the variables 
$({\cal E}_c^V , {\bf P}_c^V, \cdots)$ implies that 
they are defined using the distribution function $F_a^V$ 
instead of $F_a$.

    As explained in Ref.~18, 
the integral domain of Eq.~(\ref{dIG}) is not an arbitrary local one 
in the ${\bf X}$-space but it can be local only in 
the radial direction in order for Eq.~(\ref{dIG}) to be valid. 
   Then, if the variations 
$\delta t_E, \delta {\bf x}_E, \cdots$ in Eq.~(\ref{G}) 
are such that $\delta {\cal I} = 0$ holds 
for a spatiotemporal integral domain defined by 
$[t_0 - h / 2, t_0 + h / 2 ] \times [s_1, s_2]$ where 
$[s_1, s_2]$ represents an arbitrary spatial volume region 
sandwiched between two flux surfaces labeled by $s_1$ and $s_2$, 
then the conservation law is derived as  
\begin{eqnarray}
\label{gyro-conservation}
& & 
\left[
\left\langle 
\frac{\partial }{\partial t}\delta G_0^V ({\bf X}, t)
+ \nabla \cdot \delta {\bf G}^V ({\bf X}, t_0)
\right\rangle
\right]_{t=t_0}
\nonumber \\ 
& = &
\left[
\left\langle 
\frac{\partial }{\partial t}\delta G_0^V ({\bf X}, t)
\right\rangle
+ \frac{1}{V'}\frac{\partial}{\partial s}
\left( V' \left\langle 
\delta {\bf G}^V \cdot \nabla s 
\right\rangle \right)
\right]_{t=t_0}
\nonumber \\ & = & 0
. 
\end{eqnarray}
This is Noether's theorem for the gyrokinetic Vlasov-Poisson-Amp\`{e}re system. 
    In Eq.~(\ref{gyro-conservation}), $V' \equiv \partial V/\partial s$ represents the 
derivative of $V(s, t)$ with respect to $s$ and $V(s, t)$ denotes 
the volume enclosed by the flux surface with the label $s$ at the time $t$. 

   Using $F_a^V ({\bf Z}, t_0) = F_a ({\bf Z}, t_0)$ and 
comparing Eq.~(\ref{gkbav}) with Eq.~(\ref{gkv}),  we find  
\begin{equation}
\label{dFvdt}
\left[
\frac{\partial F_a^V ({\bf Z}, t)}{\partial t}
\right]_{t = t_0}
=
\left[
\frac{\partial F_a ({\bf Z}, t)}{\partial t}
\right]_{t = t_0}
- {\cal K}_a ({\bf Z}, t_0)
,
\end{equation}
   where ${\cal K}$ is defined by Eq.~(\ref{gkK}). 
   Let us also define $\delta G_0$ and $\delta {\bf G}$ from 
$\delta G_0^V$ and $\delta {\bf G}^V$ by replacing 
$F_a^V$ with $F_a$. 
   Then, we have 
${\bf G}^V ({\bf X}, t_0) = {\bf G} ({\bf X}, t_0)$ and 
\begin{equation}
\label{gyro-dG0dt}
\left[
\frac{\partial \delta G_0^V ({\bf X}, t)}{\partial t}
\right]_{t = t_0}
=
\left[
\frac{\partial \delta G_0 ({\bf X}, t))}{\partial t}
\right]_{t = t_0}
- \delta K_{G0} ({\bf X}, t_0)
, 
\end{equation}
   where 
\begin{eqnarray}
\label{dKG0}
\delta K_{G0}
& = & 
K_{{\cal E}c} \delta t_E - {\bf K}_{Pc} \cdot \delta {\bf x}_E
,
\nonumber \\
K_{{\cal E}c}
& = & 
\sum_a e_a \int dU \int d\mu \int d\xi \; D_a {\cal K}_a H_a
,
\nonumber \\
{\bf K}_{Pc} 
& = & 
\sum_a e_a \int dU \int d\mu \int d\xi \; D_a {\cal K}_a 
{\bf p}^c_a
.
\end{eqnarray}
   Here, ${\bf p}^c_a$ denotes the canonical 
momentum for species $a$ defined by 
\begin{equation}
\label{pcan}
{\bf p}^c_a
=
\frac{e_a}{c} {\bf A}_a^*
=
\frac{e_a}{c} {\bf A}_0 
+ m_a U {\bf b} 
. 
\end{equation}
Substituting Eq.~(\ref{gyro-dG0dt}) into Eq.~(\ref{gyro-conservation}) 
and rewriting the arbitrarily chosen time $t_0$ as $t$,   
we obtain the conservation law 
for the gyrokinetic Boltzmann-Poisson-Amp\`{e}re system, 
\begin{eqnarray}
\label{conservation}
& & 
\left\langle 
\frac{\partial }{\partial t}\delta G_0 ({\bf X}, t)
+ \nabla \cdot \delta {\bf G} ({\bf X}, t)
\right\rangle
\nonumber \\ 
& = &
\left\langle 
\frac{\partial }{\partial t}\delta G_0 ({\bf X}, t)
\right\rangle
+ \frac{1}{V'}\frac{\partial}{\partial s}
\left( V' \left\langle 
\delta {\bf G} \cdot \nabla s 
\right\rangle \right)
\nonumber \\ 
& = &
\left\langle 
\delta K_{G0}
\right\rangle
, 
\end{eqnarray}
  where 
$\left\langle \delta K_{G0} \right\rangle$ 
represents effects of the collision and source terms on the conservation law. 
     Under the nonstationary background field ${\bf B}_0$, flux surfaces may change their shapes and the grid of the flux coordinates moves. 
    Then, Eq.~(\ref{conservation}) is rewritten as 
\begin{equation}
\label{cons2}
\frac{\partial}{\partial t}
\left( V' \left\langle \delta G_0 
\right\rangle \right)
+ \frac{\partial}{\partial s}
\left( V' \left\langle 
\left( \delta {\bf G} - \delta G_0  {\bf u}_s 
\right) \cdot \nabla s
\right\rangle \right) 
= 
V'
\left\langle 
\delta K_{G0}
\right\rangle
, 
\end{equation}
   where
${\bf u}_s \cdot \nabla s$ represents 
the radial velocity of the flux surface labeled by $s$ and 
${\bf u}_s$ is defined by 
$
{\bf u}_s  
= \partial {\bf x}
(s, \theta, \zeta, t)/\partial t
$
with the flux coordinates $(s, \theta, \zeta)$ [see Eq.~(2.35) in Ref.~31]. 
     In Sec.~V, gyrokinetic energy and toroidal angular  momentum balance equations 
are derived from Eq.~(\ref{cons2}).

\section{EQUATIONS FOR GYROCENTER DENSITIES AND POLARIZATION}

   In this section, we take the velocity-space integral of the gyrokinetic 
Boltzmann equation in Eq.~(\ref{gkbav}) to consider the particle 
transport before treating the energy and toroidal angular momentum 
transport in Sec.~V. 
   We define the gyrocenter density $n_a^{\rm (gc)}$ by 
\begin{equation}
\label{ngc}
n_a^{\rm (gc)} ({\bf X}, t) 
= 
\int d U \int d\mu \int d\xi  \; 
D_a F_a 
,
\end{equation}
and the gyrocenter flux 
$\mbox{\boldmath$\Gamma$}_a^{\rm (gc)}$ 
by 
\begin{equation}
\label{ugc}
\mbox{\boldmath$\Gamma$}_a^{\rm (gc)} 
= 
n_a^{\rm (gc)} {\bf u}_a^{\rm (gc)} =
\int d U \int d\mu \int d\xi \; D_a F_a 
{\bf v}_a^{\rm (gc)}
,  
\end{equation}
where ${\bf u}_a^{\rm (gc)}$ represents 
the gyrocenter fluid velocity and 
the gyrocenter drift velocity ${\bf v}_a^{\rm (gc)} = d {\bf X}_a/dt$ is 
given by evaluating the right-hand side of Eq.~(\ref{dXdt}) at $({\bf X}, U, \mu)$. 
   Then, integrating the gyrokinetic Boltzmann equation, Eq.~(\ref{gkb1}), 
with respect to the gyrocenter velocity-space coordinates $(U, \mu, \xi)$ 
and using Eq.~(\ref{gkK}), we obtain 
\begin{equation}
\label{gcc}
\frac{\partial n_a^{\rm (gc)}}{\partial t}
+ \nabla  
\cdot 
\left( 
\mbox{\boldmath$\Gamma$}_a^{\rm (gc)} 
+ \mbox{\boldmath$\Gamma$}_a^{\rm C}
\right)
= 
\int d U \int d\mu \int d\xi \; D_a 
{\cal S}_a
.
\end{equation}
     Using the approximate collision operator given by Eq.~(\ref{DCg2}) in Appendix~C, 
the particle flux $\mbox{\boldmath$\Gamma$}_a^{\rm C}$ due to 
collisions and finite gyroradii 
is defined by Eq.~(\ref{JAC3}) with putting ${\cal A}_a^p ({\bf z}) = 1$. 
[If the collision operator given by Eq.~(\ref{scalar_C}) in Appendix~B is employed, 
$\mbox{\boldmath$\Gamma$}_a^{\rm C}$  is defined by Eq.~(\ref{GammaC}).] 

  As shown in Ref.~18, 
the gyrokinetic Poisson equation in Eq.~(\ref{gkp}) is rewritten as 
\begin{equation}
\label{gccd}
\sum_a e_a n_a^{\rm (gc)}
= 
 \nabla \cdot 
\left( \frac{{\bf E}_L}{4 \pi}
+ {\bf P}^{\rm (pol)}
\right),  
\end{equation}
   where ${\bf E}_L = - \nabla \phi$, 
$\nabla = \partial/\partial {\bf X}$, and 
${\bf P}^{\rm (pol)}$ represents the polarization density defined by 
\begin{eqnarray}
{\bf P}^{\rm (pol)}
 & = & 
\sum_a e_a \sum_{n=1}^\infty 
\frac{(-1)^{n-1}}{n!}
\sum_{i_1, \cdots, i_n}
 \int dU \int d\mu \int d\xi 
\nonumber \\ & & \mbox{} \times
\frac{\partial^{n-1} ( D_a F_a^*
\mbox{\boldmath$\rho$}_a \rho_{a i_1} \cdots 
\rho_{a i_{n-1}}
)
}{\partial X_{i_1} \cdots \partial X_{i_{n-1}}}
. 
\end{eqnarray}
   Here, $\rho_{a i}$ denotes the $i$th Cartesian component of $\mbox{\boldmath$\rho$}_a = {\bf b}({\bf X}, t) \times {\bf v}_{a0}({\bf Z}, t) / \Omega_a({\bf X}, t)$, and  
$
F_a^* = F_a + 
(e_a \widetilde{\psi}_a/B_0)
(\partial F_a / \partial \mu)
$.

As mentioned before Eq.~(\ref{B0}) in Sec.~III,  
$\sum_a e_a \int dU \int d\mu \int d\xi \; D_a {\cal S}_a ({\bf Z}, t) = 0$ 
is assumed. 
   Then, using Eq.~(\ref{gcc}), we can obtain the charge conservation law, 
\begin{equation}
\label{gccc}
\frac{\partial }{\partial t}
\left( \sum_a e_a n_a^{\rm (gc)} \right)
+ \nabla \cdot 
( {\bf j}^{\rm (gc)} + {\bf j}^{\rm C} )
=
0, 
\end{equation}
where the current density due to the gyrocenter drift and that due to  
the collisional particle transport are given by  
$
{\bf j}^{\rm (gc)} = \sum_a e_a 
\mbox{\boldmath$\Gamma$}_a^{\rm (gc)} 
= \sum_a e_a n_a^{\rm (gc)} {\bf u}_a^{\rm (gc)}
$
and 
$
{\bf j}^{\rm C} = \sum_a e_a \mbox{\boldmath$\Gamma$}_a^{\rm C}
$, 
respectively. 
   Note that the magnetization current is solenoidal 
and, accordingly, it does not 
contribute to the charge conservation law in Eq.~(\ref{gccc}). 
    Equation~(\ref{gccd}) is substituted into Eq.~(\ref{gccc}) to show 
\begin{equation}
\label{jgcl}
{\bf j}^{\rm (gc)}_L + {\bf j}^{\rm C}_L
= 
- \frac{\partial }{\partial t} 
\left( \frac{{\bf E}_L}{4 \pi}
+ {\bf P}^{\rm (pol)}_L
\right)
,
\end{equation}
   where the subscript $L$ is used to represent 
the longitudinal part of the vector variable. 
   Then, using Eqs.~(\ref{gccd}), (\ref{gccc}) and (\ref{jgcl}), 
we find that the useful formula,
\begin{eqnarray}
\label{calA}
& & 
\left\langle
 \frac{\partial }{\partial t}
\left( {\cal A}
\sum_a e_a n_a^{\rm (gc)} \right) 
\right\rangle
+ \left\langle
\nabla \cdot 
\left( {\cal A} {\bf j}^{\rm (gc)}_L \right)
\right\rangle
\nonumber \\ 
&  & =
\left\langle
 \frac{\partial {\cal A} }{\partial t} 
\sum_a e_a n_a^{\rm (gc)} 
\right\rangle
+ \left\langle
{\bf j}^{\rm (gc)}_L \cdot 
\nabla {\cal A} 
\right\rangle
- \left\langle
{\cal A} \left( \nabla \cdot {\bf j}^{\rm C}_L \right)
\right\rangle
\nonumber \\ 
&  & =
\left\langle
\nabla \cdot \left[
 \frac{\partial {\cal A} }{\partial t} 
\left( \frac{{\bf E}_L}{4 \pi}
+ {\bf P}^{\rm (pol)}_L 
\right) 
-  {\cal A} {\bf j}^{\rm C}_L
\right] \right\rangle
\nonumber \\ & & \mbox{}
\hspace*{5mm}
- \left\langle
\frac{\partial}{\partial t} 
\left[
\left( \frac{{\bf E}_L}{4 \pi}
+ {\bf P}^{\rm (pol)}_L
\right) 
\cdot 
\nabla {\cal A} 
\right\rangle \right]
,
\end{eqnarray}
holds for any function ${\cal A}({\bf X}, t)$. 
   The relation in Eq.~(\ref{calA}) is used in Sec.~V.B to derive Eq.~(\ref{tamc1}).

\section{GYROKINETIC ENERGY AND TOROIDAL ANGULAR 
MOMENTUM BALANCE EQUATIONS}

In this section, energy and toroidal angular momentum balance equations 
for gyrokinetic systems including collisional processes are derived by using the results 
shown in Secs.~III--IV.

\subsection{Energy balance equation}

The variation $\delta {\cal I}$ of the action given in Sec.~III vanishes 
under the infinitesimal time translation represented 
by $\delta t_E = \epsilon$ where $\epsilon$ is an infinitesimally small constant. 
   Here, all other infinitesimal 
variations $\delta {\bf x}_E$,  $\delta \phi$, $\cdots$ are regarded as zero. 
   Then, $\delta G_0$ and $\delta {\bf G}$ are determined by these conditions for 
the infinitesimal time translation and they satisfy Eq.~(\ref{cons2}) which, 
in the same manner as in Ref.~18, 
leads to the energy balance equation, 
\begin{equation}
\label{Econs2}
\frac{\partial}{\partial t}
\left( V' \left\langle {\cal E} 
\right\rangle \right)
+ \frac{\partial}{\partial s}
\left( V' \left\langle 
\left( {\bf Q}_c^* + {\bf Q}_R^* - {\cal E} {\bf u}_s 
\right) \cdot \nabla s
\right\rangle \right) 
= 
V' 
\left\langle 
K_{{\cal E}c}
\right\rangle
. 
\end{equation}
   Here,  the energy density ${\cal E}$ is defined by 
\begin{eqnarray}
\label{E=}
{\cal E} & = & 
{\cal E}_c
+ \nabla \cdot 
\left( \frac{1}{4\pi} \phi \nabla \phi -
\mbox{\boldmath$\Phi$}_R \right) 
\nonumber \\ 
& = & 
\sum_a \int dU \int d\mu \int d\xi \; 
D_a F_a 
\left(
\frac{m_a}{2}
\left|  {\bf v}_{a0} - \frac{e_a}{m_a c}{\bf A}_1
\right|^2
\right.
\nonumber \\ & & 
\left. \mbox{}
+ \frac{e_a}{2 B_0} \frac{\partial}{\partial \mu}
\left\langle \widetilde{\psi}_a
\left( 2 \widetilde{\phi} - \widetilde{\psi}_a
\right) \right\rangle_\xi
\right)
\nonumber \\ & & \mbox{}
+ \frac{1}{8\pi} 
\left( |\nabla \phi|^2 
+ |{\bf B}_0 + {\bf B}_1 |^2 \right)
,
\end{eqnarray}
   and the energy fluxes ${\bf Q}_c^*$ and ${\bf Q}_R^*$ 
are given by 
\begin{eqnarray}
\label{Qc*}
{\bf Q}_c^*  & = & 
{\bf Q}_c - \frac{1}{4\pi} \frac{\partial}{\partial t}
\left(  \phi \nabla \phi \right) 
\nonumber \\ & = & \mbox{}
\sum_a \int dU \int d\mu \int d\xi \; 
D_a F_a \left[ H_a {\bf v}_a^{\rm (gc)}
+ \frac{\partial {\bf A}_0}{\partial t}
\right.
\nonumber \\ &  & \mbox{}
\left.
\times \left(
-\mu {\bf b} + \frac{m_a U}{B_0}
({\bf v}_a^{\rm (gc)})_\perp
- {\bf N}_a
\right) 
\right]
- \frac{1}{4\pi} \phi
 \nabla \frac{\partial \phi}{\partial t}
\nonumber \\ & & \mbox{}
- \frac{1}{4\pi}
\frac{\partial ({\bf A}_0 + {\bf A}_1 )}{\partial t}
\times ( {\bf B}_0 + {\bf B}_1 )
+ \frac{1}{4\pi c}
\left(
\lambda \frac{\partial {\bf A}_1}{\partial t}
\right.
\nonumber \\ & & 
\left. \mbox{}
+ \alpha \frac{\partial {\bf A}_0}{\partial t}
\right)
- \frac{1}{4\pi}\mbox{\boldmath$\Lambda$} \times 
\left(
\frac{\partial {\bf A}_0}{\partial t}
+ \frac{\partial \chi}{\partial t} \nabla \zeta
\right)
\end{eqnarray}
   and 
\begin{eqnarray}
\label{QR*}
{\bf Q}_R^* & = & 
{\bf Q}_R + 
\frac{\partial \mbox{\boldmath$\Phi$}_R}{\partial t}
,
\end{eqnarray}
   respectively, where ${\bf N}_a$, 
${\bf Q}_R$, and $\mbox{\boldmath$\Phi$}_R$ are 
defined by Eqs.~(43), (85), and (88) in Ref.~18, respectively. 

      Recalling Eq.~(\ref{dKG0}) and 
using Eq.~(\ref{DCgA3}) into which we substitute ${\cal A}_a^g ({\bf Z})= H_a ({\bf Z})$  and the approximate collision operator 
given by Eq.~(\ref{DCg2}) 
in Appendix~C, 
we rewrite 
the right-hand side of Eq.~(\ref{Econs2}) as 
\begin{eqnarray}
\label{QCS}
V' 
\left\langle 
K_{{\cal E}c}
\right\rangle
& = & 
-
\frac{\partial}{\partial s}
\left( V' \left\langle 
{\bf Q}^{\rm C}
\cdot \nabla s
\right\rangle \right)
\nonumber \\ & & 
\mbox{}
+ 
V' 
\sum_a
\left\langle 
\int d U \int d\mu \int d\xi \; D_a 
{\cal S}_a H_a
\right\rangle
,
\nonumber \\ & & 
\end{eqnarray}
   where the energy flux ${\bf Q}^{\rm C}$ due to collisions and finite gyroradii 
is defined by taking the summation of Eq.~(\ref{JAC3}) over species $a$ with 
putting ${\cal A}_a^p ({\bf z}) = \frac{1}{2}m_a v_{\parallel a}^2 
+ \mu_{0a} B_0 ({\bf x}_a) + e_a \phi ({\bf x}_a)$. 
   The derivation of Eq.~(\ref{QCS}) requires 
Eq.~(\ref{CHp=0}) which is satisfied by the 
collision operator in the form of Eq.(\ref{DCg2}) 
with appropriately choosing $\Delta {\bf x}_a^{(2)}$, 
$\Delta v_{\parallel a}$, and $\Delta \mu_{0 a}$ 
as described in Appendix~C. 

Substituting Eq.~(\ref{QCS}) into Eq.~(\ref{Econs2}), 
the energy balance equation is rewritten as
\begin{eqnarray}
\label{Econs2a}
& & 
\frac{\partial}{\partial t}
\left( V' \left\langle {\cal E} 
\right\rangle \right)
+ \frac{\partial}{\partial s}
\left( V' \left\langle 
\left( {\bf Q} - {\cal E} {\bf u}_s 
\right) \cdot \nabla s
\right\rangle \right) 
\nonumber \\ & & 
= 
V' 
\sum_a
\left\langle 
\int d U \int d\mu \int d\xi \; D_a 
{\cal S}_a H_a
\right\rangle
, 
\end{eqnarray}
where the energy flux ${\bf Q}$ is given by 
\begin{eqnarray}
\label{Q}
{\bf Q} & = & 
{\bf Q}_c^* + {\bf Q}_R^* + {\bf Q}^{\rm C}
. 
\end{eqnarray}
    The right-hand side of Eq.~(\ref{Econs2a}) represents the external energy source. 
  It is confirmed later in Sec.~VI.B that the ensemble average of 
$\langle {\bf Q} \cdot \nabla s \rangle $ coincides 
with the well-known expression of the radial energy transport to the lowest order 
in the $\delta$-expansion.

\subsection{Toroidal angular momentum balance equation}

The toroidal angular momentum balance equation is derived from the fact that 
$\delta {\cal I} = 0$ under the infinitesimal toroidal rotation represented by 
$\delta {\bf x}_E = \epsilon {\bf e}_\zeta ({\bf X})$. 
Here, $\epsilon$ is again an infinitesimally small constant, and ${\bf e}_\zeta({\bf X})$ is defined by   
$
{\bf e}_\zeta ({\bf X})
= 
\partial {\bf X}(R, z, \zeta)/\partial \zeta
= R^2 \nabla \zeta
$  
where the right-handed cylindrical spatial coordinates $(R, z, \zeta)$ are used. 
   We also define $\hat{\bf z}$ by 
$
\hat{\bf z} = R\nabla \zeta \times \nabla R
$ 
which represents the unit vector in the $z$-direction. 
  Then, if putting the origin of the position vector ${\bf X}$ at $(R, z) = (0, 0)$, 
we have 
$
{\bf e}_\zeta ({\bf X}) =  {\bf X} \times \hat{\bf z}
$.
  Under the infinitesimal toroidal rotation, the variations of the vector variables 
are given as 
$
\delta {\bf A}_1 = \epsilon {\bf A}_1 \times \hat{\bf z} 
$
and
$
\delta {\bf A}_0 = \epsilon {\bf A}_0 \times \hat{\bf z} 
$
although the other variations
$\delta t_E$, $\delta \phi$, $\cdots$, are all regarded as zero.  
   Then, using these variations of the variables 
associated with the infinitesimal toroidal rotation, 
the canonical momentum balance equation is derived from 
Eq.~(\ref{conservation}) as 
\begin{eqnarray}
\label{tamc0}
& & 
\left\langle
\frac{\partial (
{\bf P}_c \cdot 
{\bf e}_\zeta
)
}{\partial t} 
\right\rangle
+ 
\frac{1}{V'} \frac{\partial}{\partial s}
\left[ V'
\left\langle
\nabla s \cdot
\left(
\mbox{\boldmath$\Pi$}_c \cdot {\bf e}_\zeta
\right. \right. \right.
\nonumber \\ & & 
\left. \left. \left. \mbox{}
+ \left(
\mbox{\boldmath$\Sigma$}_{A1} \times {\bf A}_1
+ 
\mbox{\boldmath$\Sigma$}_{A0} \times {\bf A}_0
\right) \cdot \hat{\bf z}
+ {\bf P}_{R\zeta} 
\right)
\right\rangle
\right]
\nonumber \\ & & 
= 
\left\langle K_{Pc\zeta} \right\rangle
.
\end{eqnarray}
     Here, the density of the canonical toroidal angular momentum is defined by  
\begin{equation}
\label{Pcez}
{\bf P}_c \cdot 
{\bf e}_\zeta
=
\sum_a \int dU \int d\mu \int d\xi \; D_a F_a
(p_a^c)_\zeta
, 
\end{equation}
with the toroidal component of the canonical momentum 
denoted by
\begin{equation}
\label{pcz}
(p_a^c)_\zeta
=
\frac{e_a}{c} A_{a \zeta}^*
=
\frac{e_a}{c} A_{0 \zeta}
+ m_a U b_\zeta
, 
\end{equation}
where $b_\zeta \equiv I/B_0$ represents the covariant toroidal component of 
${\bf b} \equiv {\bf B}_0 / B_0$.  
   Definitions of $\mbox{\boldmath$\Sigma$}_{A1}$, 
$\mbox{\boldmath$\Sigma$}_{A0}$, and  
${\bf P}_{R\zeta}$ on the left-hand side of Eq.~(\ref{tamc0}) 
are given by Eqs.~(79) and (100) in Ref.~18. 
     On the right-hand side of Eq.~(\ref{tamc0}), 
the variation of the canonical toroidal angular momentum
due to collisions and external sources is given by 
\begin{eqnarray}
\label{Kpcz}
& & 
K_{Pc\zeta}
=
{\bf K}_{Pc} \cdot {\bf e}_\zeta
= 
\sum_a \int dU \int d\mu \int d\xi \; D_a 
{\cal K}_a 
(p_a^c)_\zeta
.
\nonumber \\ & & 
\end{eqnarray}

   We  follow the same procedures as shown in 
Sec.~V.B of Ref.~18 and use 
Eqs.~(\ref{tamc0})--(\ref{Kpcz}) and Eq.~(\ref{calA}) 
with ${\cal A} = A_{0\zeta} = - \chi$
to write the toroidal angular momentum balance equation as 
\begin{eqnarray}
\label{tamc1}
& & 
\left\langle 
\frac{\partial}{\partial t}
\left[
P_{\parallel \zeta}
- \frac{1}{c} \left( {\bf P}^{\rm (pol)}_L 
+ \frac{{\bf E}_L}{4\pi} \right) 
\cdot \nabla A_{0\zeta} 
\right] \right\rangle
\nonumber \\ 
& & +
\frac{1}{V'}\frac{\partial}{\partial s}
\left[ V' \left\{ \Pi_{\parallel\zeta}^s 
+ \Pi_{R\zeta}^s 
- \frac{1}{4\pi}
\left\langle A_{1\zeta} (\nabla \times {\bf B}_1) \cdot \nabla s 
\right\rangle
\right. \right. \nonumber 
\\ & & \mbox{}
- \frac{1}{4\pi}
\left\langle E_{L\zeta} E_L^s + B_{1\zeta}B_1^s 
\right\rangle
+ \frac{1}{4\pi c}
\left\langle \frac{\partial \lambda}{\partial \zeta} A_1^s 
\right\rangle
\nonumber \\ & & 
\left. \left. \mbox{}
+ \frac{1}{c}  \left\langle 
\frac{\partial A_{0\zeta}}{\partial t}
\left( {\bf P}^{\rm (pol)}_L 
+ \frac{{\bf E}_L}{4\pi} \right)\cdot \nabla s 
\right\rangle
\right\} \right] 
\nonumber \\ & & 
=
\left\langle K_{Pc\zeta} \right\rangle 
+ \frac{1}{c}
\left\langle  \nabla \cdot  
\left( A_{0\zeta} {\bf j}_L^{\rm C} \right)
 \right\rangle
, 
\end{eqnarray}
     where 
\begin{eqnarray}
\label{Pparaz}
& & 
P_{\parallel\zeta}
 = \sum_a \int dU \int d\mu \int d\xi \; D_a F_a 
m_a U b_\zeta 
, 
\nonumber \\ & & 
\Pi_{\parallel\zeta}^s
=
\sum_a \int dU \int d\mu \int d\xi \; D_a F_a 
m_a U b_\zeta 
{\bf v}_a^{\rm (gc)} \cdot \nabla s
, 
\nonumber \\ & & 
\Pi_{R\zeta}^s
 = {\bf P}_{R\zeta} \cdot \nabla s
.
\end{eqnarray}
   Using Eqs.~(\ref{Kpcz}) and (\ref{DCgA3}) in Appendix~C 
with putting ${\cal A}_a^g = (p_a^c)_\zeta$, 
the right-hand side of Eq.~(\ref{tamc1}) is rewritten as 
\begin{eqnarray}
\label{PiCS}
& & 
\left\langle K_{Pc\zeta} \right\rangle 
+ \frac{1}{c}
\left\langle  \nabla \cdot  
\left( A_{0\zeta} {\bf j}_L^{\rm C} \right)
 \right\rangle
\nonumber \\
& = & 
-
\frac{1}{V'}
\frac{\partial}{\partial s}
\left[ V' \left\langle 
\left(
{\bf J}_{p\zeta}^{\rm C}
+ \frac{\chi}{c} {{\bf j}_L^{\rm C} }
\right)
\cdot \nabla s
\right\rangle \right]
\nonumber \\ & & 
\mbox{}
+ 
\sum_a
\left\langle 
\int d U \int d\mu \int d\xi \; D_a 
{\cal S}_a m_a U b_\zeta 
\right\rangle
,
\end{eqnarray}
   where ${\bf J}_{p\zeta}^{\rm C}$ is defined by taking 
the summation of Eq.~(\ref{JAC3}) over species $a$ with 
putting ${\cal A}_a^p ({\bf z}) = (e_a/c) [ A_{0\zeta}({\bf x}) 
+ A_{1\zeta}({\bf x}) ] + m_a v_\zeta$. 
   In deriving Eq.~(\ref{PiCS}), 
$\sum_a e_a \int dU \int d\mu \int d\xi \; D_a {\cal S}_a ({\bf Z}, t) = 0$ 
and Eq.~(\ref{Cpp=0}) are used. 
    The approximate collision operator which satisfies Eq.~(\ref{Cpp=0}) 
is presented in Appendix~C.

Substituting Eq.~(\ref{PiCS}) into Eq.~(\ref{tamc1}), 
the toroidal angular momentum balance equation is rewritten as 
\begin{eqnarray}
\label{tamc1a}
& & 
\frac{\partial }{\partial t}
\left( V' 
\left\langle 
P_{\parallel \zeta}
- \frac{1}{c} \left( {\bf P}^{\rm (pol)}_L 
+ \frac{{\bf E}_L}{4\pi} \right) 
\cdot \nabla A_{0\zeta} 
 \right\rangle
\right)
\nonumber \\ & & 
 +
\frac{1}{V'}\frac{\partial}{\partial s}
\left[ V' \left\{ \Pi_{\parallel\zeta}^s 
+ \Pi_{R\zeta}^s +  (\Pi^{\rm C})^s
- \frac{1}{4\pi}
\right. \right. \nonumber 
\\ & & \mbox{}
\times
\left\langle A_{1\zeta} (\nabla \times {\bf B}_1) \cdot \nabla s 
\right\rangle
- \frac{1}{4\pi}
\left\langle E_{L\zeta} E_L^s + B_{1\zeta}B_1^s 
\right\rangle
\nonumber \\ & & 
\mbox{}
+ \frac{1}{4\pi c}
\left\langle \frac{\partial \lambda}{\partial \zeta} A_1^s 
\right\rangle
+ \frac{1}{c}  \left\langle 
\frac{\partial A_{0\zeta}}{\partial t}
\left( {\bf P}^{\rm (pol)}_L 
+ \frac{{\bf E}_L}{4\pi} \right)\cdot \nabla s 
\right\rangle
\nonumber \\ & & 
\left. \left. \mbox{}
 - 
\left\langle \left[ 
P_{\parallel \zeta}
- \frac{1}{c} \left( {\bf P}^{\rm (pol)}_L 
+ \frac{{\bf E}_L}{4\pi} \right) 
\cdot \nabla A_{0\zeta} 
\right] 
( {\bf u}_s \cdot \nabla s ) \right\rangle 
\right\} \right] 
\nonumber \\ & & 
=  
\sum_a
\left\langle 
\int d U \int d\mu \int d\xi \; D_a 
{\cal S}_a m_a U b_\zeta 
\right\rangle
, 
\end{eqnarray}
where the right-hand side represents the external source of the toroidal angular momentum 
and 
\begin{equation}
\label{PiC}
(\Pi^{\rm C})^s = 
\left(
{\bf J}_{p\zeta}^{\rm C}
+ \frac{\chi}{c} {{\bf j}_L^{\rm C} }
\right)
\cdot \nabla s
\end{equation}
is the radial flux of the toroidal angular momentum due to 
collisions and finite gyroradii. 
   In Sec.~VI.C, we derive the ensemble-averaged toroidal 
angular momentum balance equation from Eq.~(\ref{tamc1a}) 
in order to confirm that it is consistent with the conventional result 
up to the second order in $\delta$. 

\section{ENSEMBLE-AVERAGED BALANCE EQUATIONS FOR PARTICLES, ENERGY, AND TOROIDAL ANGULAR MOMENTUM}

In this section, the particle, energy and toroidal angular momentum balance equations 
derived in Secs.~IV and V are ensemble-averaged for the purpose of verifying 
their consistency with those obtained by conventional recursive  
formulations.~\cite{Sugama1996,Sugama1997,Sugama1998,Sugama2011}
    In the same way as shown in Sec.~VI of Ref.~18, 
we divide an arbitrary physical variable 
${\cal Q}$ into the average and turbulent parts as 
\begin{equation}
\label{QQQ}
{\cal Q} = \langle Q \rangle_{\rm ens} 
+ \hat{\cal Q} 
, 
\end{equation}
    where $\langle \cdots \rangle_{\rm ens}$ represents the ensemble average, and  
    we immediately find 
$
\langle \hat{\cal Q}  \rangle_{\rm ens}   = 0
$.
    We identify the zeroth fields ${\bf A}_0$ and ${\bf B}_0$ 
with the ensemble-averaged parts to write  
$
{\bf A}_0  
 =  
\langle {\bf A} \rangle_{\rm ens}, 
$
$
{\bf A}_1  
 =  
\hat{\bf A}, 
$
$
{\bf B}_0  
 =  
\langle {\bf B} \rangle_{\rm ens}, 
$
and
$
{\bf B}_1  
 =  
\hat{\bf B}
$.
   Regarding the electrostatic potential $\phi$, 
it is written as the sum of the average and fluctuation parts, 
$
\phi ({\bf x}, t) 
=
\langle \phi ({\bf x}, t ) \rangle_{\rm ens} 
+ \hat{\phi}  ({\bf x}, t)
.
$
    Here, assuming that $\langle \phi ({\bf x}, t ) \rangle_{\rm ens} \neq 0$, 
the background ${\bf E} \times {\bf B}$ flow is retained and its velocity is regarded 
as ${\cal O}(\delta v_T)$, where $\delta$ and $v_T$ represent the drift ordering parameter and the thermal velocity, respectively. 
    Then, using Eq.~(\ref{psi=}), we have 
$
\psi_a 
=
\langle \psi_a \rangle_{\rm ens}
+ \hat{\psi}_a, 
$
   where 
\begin{equation}
\label{psia}
\langle \psi_a \rangle_{\rm ens}
= \langle \phi \rangle_{\rm ens}, 
\hspace*{5mm}
\hat{\psi}_a
=
\hat{\phi} - \frac{{\bf v}_0}{c} \cdot 
\hat{\bf A} 
.
\end{equation}

   We assume that the ensemble average 
$\langle {\cal Q} \rangle_{\rm ens}$ of any variable ${\cal Q}$ considered here has a slow temporal variation subject to the so-called transport ordering, 
$
\partial
\ln \langle {\cal Q} \rangle_{\rm ens} / \partial t 
=
{\cal O} (\delta^2 v_T / L)
$,
and that it has a gradient scale length $L$ which is on the same order as gradient scale lengths of the equilibrium field and pressure profiles. 
    We also impose the constraint of axisymmetry on 
$\langle {\cal Q} \rangle_{\rm ens}$ that is written as 
$
\partial \langle {\cal Q} \rangle_{\rm ens} / \partial \zeta
= 0 
$
even though ${\cal Q}$ itself is not axisymmetric. 
    On the other hand, the turbulent part $\hat{\cal Q}$ of 
${\cal Q}$ is assumed to  vary with a characteristic 
frequency  $\omega = {\cal O} (v_T/L)$ and 
have gradient scale lengths  $L$ and $\rho$ in the directions 
parallel and perpendicular to the equilibrium 
magnetic field ${\bf B}_0$, respectively.

The ensemble-averaged part $\langle F_a \rangle_{\rm ens}$ of 
the distribution function $F_a$ for species $a$ consists of 
the local Maxwellian part and the deviation from it, 
\begin{equation}
\label{Fens=}
\langle F_a \rangle_{\rm ens}
 = F_{aM} 
+ \langle F_{a1} \rangle_{\rm ens}
.  
\end{equation}
       The local Maxwellian distribution function is written as 
$
F_{aM}  = n_{a0} [m_a/(2\pi T_{a0}) ]^{3/2}
\exp [ - 
( m_a U^2 / 2 + \mu B_0 ) / T_{a0} ]
$
    where the equilibrium density $n_{a0}$ and temperature $T_{a0}$ are regarded as uniform on flux surfaces.  
    The first-order ensemble-averaged distribution function $\langle F_{a1} \rangle_{\rm ens}$ is determined by the drift kinetic equation, which can be derived by substituting Eq.~(\ref{Fens=}) into the ensemble average of Eq.~(\ref{gkb}).  
    The derived equation agrees, to ${\cal O}(\delta)$, with the well-known linearized drift kinetic equation, on which the neoclassical transport theory 
is based.~\cite{H&S,Helander} 

      The fluctuation part $\hat{F}_a$ is written as 
\begin{equation}
\label{hatF}
 \hat{F}_a = - F_{aM} \frac{e_a \langle \hat{\psi}_a \rangle_\xi}{T_a} + \hat{h}_a
.
\end{equation}
   Substituting Eq.~(\ref{hatF})
into the fluctuation part of the gyrokinetic equation in Eq.~(\ref{gkb}) 
yields 
\begin{eqnarray}
\label{hgke}
& & \frac{\partial \hat{h}_a}{\partial t}
+ \{ \hat{h}_a, H_a \} 
\nonumber \\ 
& = & 
F_{aM} \left[ 
\frac{e_a}{T_{a0}}
\frac{\partial \langle \hat{\psi}_a \rangle_\xi}{\partial t}
- \hat{\bf v}_a^{\rm (gc)} \cdot
\left( \nabla \ln p_{a0}
+ \frac{e_a}{T_{a0}} 
\nabla \langle \phi \rangle_{\rm ens} 
\right. \right. 
\nonumber \\ & & 
\left. \left. \mbox{}
+ \left(\frac{\frac{1}{2}m_a U^2 + \mu B_0 }{T_{a0}} - \frac{5}{2}
\right)
\nabla \ln T_{a0} 
\right)
\right]
+ C_a^L 
\end{eqnarray}
   where $C_a^L$ represents the linear collision term defined by
\begin{eqnarray}
C_a^L ({\bf X})
& = & 
\sum_b
\langle
[ 
C_{ab}^p ( \hat{h}_a ({\bf x} - \mbox{\boldmath$\rho$}_a ),  F_{bM} )
\nonumber \\ & & 
\mbox{}
+ C_{ab}^p ( F_{aM}, \hat{h}_b ({\bf x} - \mbox{\boldmath$\rho$}_b ) ) 
]_{{\bf x} = {\bf X} + \mbox{\boldmath$\rho$}_a}
\rangle_{\xi_a}
. 
\hspace*{5mm}
\end{eqnarray}
   Equation~(\ref{hgke}) is valid to the lowest order in $\delta$ and 
agrees with the conventional gyrokinetic equation for 
the nonadiabatic part $\hat{h}_a$ of the perturbed distribution function 
derived from using the 
WKB representation.~\cite{Frieman,Sugama1996}  
   On the right-hand side of Eq.~(\ref{hgke}), the turbulent part $\hat{\bf v}_a^{\rm (gc)}$ of the gyrocenter drift velocity 
${\bf v}_a^{\rm (gc)} = d {\bf X}_a / dt = \{ {\bf X}_a, H_a \}$
is written as   
\begin{equation}
\label{vgc_turb}
\hat{\bf v}_a^{\rm (gc)}  = 
\frac{c}{B_0} {\bf b}
\times  \nabla \langle \hat{\psi}_a 
({\bf X} + \mbox{\boldmath$\rho$}_a, t)  \rangle_\xi 
+ {\cal O}(\delta^2) 
. 
\end{equation}
   It is shown by using Eq.~(\ref{hatF}) and the WKB representation 
that, to the lowest order in $\delta$, 
the turbulent parts of Eqs.~(\ref{gkp}) and (\ref{ampere1b}) 
agree with the gyrokinetic Poisson equation and the gyrokinetic Amp\`{e}re's law 
derived by conventional recursive formulations.~\cite{AL,Sugama1996}

\subsection{Ensemble-averaged particle balance equation}

   Taking the ensemble average of Eq.~(\ref{gcc}) and 
subsequently its flux surface average, we obtain 
\begin{eqnarray}
\label{gcc2}
& & \left\langle 
 \frac{\partial \langle n_a^{\rm (gc)} \rangle_{\rm ens}}{\partial t}
\right\rangle
+ \frac{1}{V'} \frac{\partial}{\partial s} 
\left( V'  \left\langle \left\langle
 \left( 
\mbox{\boldmath$\Gamma$}_a^{\rm (gc)}
+  \mbox{\boldmath$\Gamma$}_a^{\rm C}
\right)
\cdot \nabla s 
\right\rangle \right\rangle  \right)
\nonumber  \\ 
& & 
= 
\left\langle
\int d U \int d\mu \int d\xi \; D_a 
{\cal S}_a
\right\rangle
,  
\end{eqnarray}
   where 
$
\langle n_a^{\rm (gc)} \rangle_{\rm ens}
= n_{a0} + {\cal O}(\delta)
$, 
   and $\langle \langle \cdots \rangle \rangle$ represents a double average over the flux surface and the ensemble. 
   Here, $n_{a0}$ is the equilibrium density which is a flux-surface function and characterizes the Maxwellian distribution function $F_{aM}$.  
  On the right-hand side, the source term ${\cal S}_a$ is regarded as of 
${\cal O} (\delta^2)$ as well as all other terms in Eq.~(\ref{gcc2}), 
and it is assumed to have no turbulent component so that 
${\cal S}_a = \langle {\cal S}_a \rangle_{\rm ens}$. 
   
   It is shown in Ref.~18 that the radial gyrocenter particle flux is given by 
\begin{equation}
\label{Gamma(gc)}
(\Gamma_a^{\rm (gc)})^s 
  \equiv
\left\langle \left\langle
\mbox{\boldmath$\Gamma$}_a^{\rm (gc)}
\cdot \nabla s 
\right\rangle \right\rangle
 =  (\Gamma_a^{\rm NA})^s + (\Gamma_a^{\rm A})^s
,
\end{equation}
   where the nonturbulent part $(\Gamma_a^{\rm NA})^s$ 
and  the turbulence-driven part $(\Gamma_a^{\rm A})^s$ 
are written as 
\begin{eqnarray}
\label{GammaNA}
& & 
(\Gamma_a^{\rm NA})^s  
\equiv
\left\langle \int dU \int d\mu \int d\xi \;
D_a \langle F_a \rangle_{\rm ens}
\langle {\bf v}_a^{\rm (gc)}
\rangle_{\rm ens}
\cdot \nabla s 
\right\rangle
\nonumber \\ 
& & = 
\left\langle
\frac{c}{e_a B_0}  
\left[ {\bf b} \times  
\left( \nabla \cdot {\bf P}_{a1}^{\rm CGL} \right) 
\right] \cdot \nabla s \right\rangle
\nonumber \\ & & \mbox{} 
\hspace*{5mm}
+ n_{a0} \left\langle
\frac{c}{B_0} 
\left( \langle {\bf E} \rangle_{\rm ens} \times  {\bf b} \right) 
 \cdot \nabla s 
\right\rangle
+ {\cal O}(\delta^3)
, 
\end{eqnarray}
  and  
\begin{eqnarray}
\label{GammaA}
& & (\Gamma_a^{\rm A})^s 
\equiv
\left\langle \int dU \int d\mu \int d\xi \;
D_a \langle \hat{F}_a
\hat {\bf v}_a^{\rm (gc)}
\rangle_{\rm ens}
\cdot \nabla s 
\right\rangle
\nonumber \\ 
& & 
=
- 
\left\langle  \left\langle
\frac{c}{B_0} 
\int dU \int d\mu \int d\xi \; D_a
\hat{h}_a
( \nabla \hat{\psi}_a \times {\bf b} ) 
\cdot \nabla s
\right\rangle \right\rangle
\nonumber \\ & & \mbox{}
+ {\cal O}(\delta^3) 
, 
\end{eqnarray}
   respectively. 
   On the right-hand side of Eq.~(\ref{GammaNA}), 
${\bf P}_{a1}^{\rm CGL}$ represents the first-order part of the pressure tensor in the Chew-Goldberger-Low (CGL) form~\cite{H&S} defined by 
$
{\bf P}_{a1}^{\rm CGL} 
=
\int dU \int d\mu \int d\xi \: D_a \langle F_{a1} \rangle_{\rm ens} \:
\left[ m_a U^2 {\bf b} {\bf b} + \mu B_0 
\left( {\bf I} - {\bf b} {\bf b} \right) \right]
$,
   and the ensemble-averaged electric field is given by  
$
\langle {\bf E} \rangle_{\rm ens} = - \nabla \langle \phi \rangle_{\rm ens}
- c^{-1} \partial {\bf A}_0/\partial t
$. 
    Thus,  Eq.~(\ref{GammaNA}) expresses 
the neoclassical radial particle flux and the radial ${\bf E} \times {\bf B}$ drift which 
are well-known by the neoclassical transport theory.~\cite{H&S} 
   We also find that Eq.~(\ref{GammaA}) agrees with the turbulent radial particle 
flux derived 
from the conventional gyrokinetic theory based on the WKB formalism.~\cite{Sugama1996}

    The radial classical particle flux is given by  
\begin{eqnarray}
\label{GammaC2}
(\Gamma_a^{\rm C})^s 
  & \equiv  & 
\left\langle \left\langle
\mbox{\boldmath$\Gamma$}_a^{\rm C}
\cdot \nabla s 
\right\rangle \right\rangle
\nonumber \\ 
& = & 
\sum_b
\left\langle \frac{m_a c}{e_a B_0}
\int dU \int d\mu \int d\xi \;
D_a 
[({\bf v} \times {\bf b} )\cdot \nabla s]
\right. 
\nonumber \\ & & 
\left.
\mbox{}
\times 
[
C_{ab}^p (\widetilde{f}_{a1}, F_{bM})
+ 
C_{ab}^p (F_{aM}, \widetilde{f}_{b1})
]
\right\rangle 
+ {\cal O}(\delta^3)
\nonumber \\ 
& = & 
\left\langle \frac{c}{e_a B_0}
( {\bf F}_{a1} \times {\bf b} )
\cdot \nabla s
\right\rangle 
+  {\cal O}(\delta^3)
,
\end{eqnarray}
%
   where ${\bf F}_{a1} \equiv 
\int d^3 {\bf v} \: m_a {\bf v} \:  C_a^p$ is the 
collisional friction force.
   It is well-known that the classical transport equation relating 
$(\Gamma_a^{\rm C})^s$ to the gradient forces is immediately derived 
from Eq.~(\ref{GammaC2}) 
because the first-order gyrophase-dependent part of the particle distribution function 
in Eq.~(\ref{GammaC2}) is expressed in terms of the gradient of 
the background Maxwellian distribution 
function as $\widetilde{f}_{a1} = - \mbox{\boldmath$\rho$}_a \cdot \nabla F_{aM}$ 
with the gradient operator $\nabla$ taken for the fixed energy variable 
$\varepsilon = \frac{1}{2} m_a v^2 + e \langle \phi \rangle_{\rm ens}$.  
    
    In the same manner as in deriving Eq.~(\ref{cons2}) from Eq.~(\ref{conservation}), 
the ensemble-averaged particle transport equation can be obtained from Eq.~(\ref{gcc2}) 
as 
\begin{eqnarray}
\label{gcc3}
& & 
\frac{\partial}{\partial t}
\left( V' n_{a0} 
 \right)
+ \frac{\partial}{\partial s}
\left( V' 
\left[ (\Gamma_a)^s 
- n_{a0} 
\langle {\bf u}_s 
 \cdot \nabla s \rangle 
\right] \right) 
\nonumber \\ & & 
\hspace*{5mm}
= 
\left\langle
\int d U \int d\mu \int d\xi \; D_a 
{\cal S}_a
\right\rangle
,
\end{eqnarray}
   where the total radial particle flux is given by the sum of 
the classical, neoclassical, and turbulent parts as 
\begin{eqnarray}
(\Gamma_a)^s 
& = & 
(\Gamma_a^{\rm (gc)})^s +  (\Gamma_a^{\rm C})^s
\nonumber \\
& = &
(\Gamma_a^{\rm NA})^s  + (\Gamma_a^{\rm A})^s 
+ (\Gamma_a^{\rm C})^s 
.
\end{eqnarray}
   As shown above, the well-known expressions of 
the classical, neoclassical and turbulent particle fluxes are included in 
$(\Gamma_a^{\rm C})^s$, 
$(\Gamma_a^{\rm NA})^s$, and $(\Gamma_a^{\rm A})^s$, respectively. 
   The latter two fluxes are evaluated by   
the solutions $\langle F_{a1} \rangle_{\rm ens}$  and $\hat{h}_a$ of the 
first-order drift kinetic and gyrokinetic equations, respectively.

\subsection{Ensemble-averaged energy balance equation}

   The ensemble average of the energy density defined by Eq.~(\ref{E=}) is written as 
\begin{equation}
\label{Eens}
\langle {\cal E} \rangle_{\rm ens}
= \frac{3}{2}\sum_a n_{a0} T_{a0}
+ \frac{B_0^2}{8\pi}
+ {\cal O}(\delta)
,
\end{equation}
   where the energy density of the electric field is neglected as a small quantity of ${\cal O}(\delta^2)$. 
   It is shown in Ref.~18 that the radial components of the first two terms on the right-hand side of Eq.~(\ref{Qc*}) are double-averaged over the ensemble and the flux surface to 
give  
\begin{eqnarray}
\label{Qc*2}
& & 
\hspace*{-3mm}
\sum_a \left\langle \left\langle \int dU \int d\mu \int d\xi \; 
D_a F_a \left( H_a {\bf v}_a^{\rm (gc)}
- \mu  \frac{\partial {\bf A}_0}{\partial t}
\times {\bf b} \right) \cdot \nabla s
\right\rangle \right\rangle
\nonumber \\ & & 
 =   \sum_a \left[ 
(q^{\rm (gc)}_a)^s + \frac{5}{2} T_{a0} (\Gamma^{\rm (gc)}_a)^s
\right] + {\cal O}(\delta^3)
.
\end{eqnarray}
   Here, the radial particle 
flux $(\Gamma_a^{\rm (gc)})^s$ is given by 
Eqs.~(\ref{Gamma(gc)})--(\ref{GammaA}) 
and the radial heat flux $(q_a)^s$ is written as 
\begin{equation}
\label{qas}
(q^{\rm (gc)}_a)^s = (q_a^{\rm NA})^s + (q_a^{\rm A})^s
\end{equation}
   which consists of the nonturbulent part, 
\begin{eqnarray}
\label{qNA}
 & & 
(q_a^{\rm NA})^s  =  
\left\langle \int dU \int d\mu \int d\xi \;
D_a \langle F_{a1} \rangle_{\rm ens}
\langle {\bf v}_a^{\rm (gc)}
\rangle_{\rm ens} \cdot \nabla s 
\right.
\nonumber \\ & & 
\left. \mbox{}
\hspace{15mm}
\times
\left( \frac{1}{2} m_a U^2 + \mu B_0 
- \frac{5}{2} T_{a0} \right)
\right\rangle
\nonumber \\ 
& & 
=
T_{a0}
\left\langle
\frac{c}{e_a B_0}  
\left[ {\bf b} \times  \left( \nabla \cdot \mbox{\boldmath$\Theta$}_{a}^{\rm CGL} \right) 
\right] \cdot \nabla s \right\rangle
+ {\cal O}(\delta^3)
,
 \hspace*{5mm}
\end{eqnarray}
   and the turbulence-driven part, 
\begin{eqnarray}
\label{qA}
& & 
(q_a^{\rm A})^s =  
- 
\left\langle  \left\langle
\frac{c}{B_0} 
\int dU \int d\mu \int d\xi \; D_a
\hat{h}_a
( \nabla \hat{\psi}_a \times {\bf b} ) 
\cdot \nabla s
\right. \right.
\nonumber \\ & & 
\hspace*{15mm}
\times
\left. \left.
\left( \frac{1}{2} m_a U^2 + \mu B_0 
- \frac{5}{2} T_{a0} \right)
\right\rangle \right\rangle
+ {\cal O}(\delta^3) 
.
\hspace*{10mm}
\end{eqnarray}
   In Eq.~(\ref{qNA}), 
the heat stress tensor $\mbox{\boldmath$\Theta$}_{a}^{\rm CGL}$ is defined by
$
T_{a0} \mbox{\boldmath$\Theta$}_{a}^{\rm CGL} 
=
\int dU \int d\mu \int d\xi \: D_a \langle F_{a1} \rangle_{\rm ens} \:
( \frac{1}{2} m_a U^2 + \mu B_0 - \frac{5}{2} T_{a0} )
[ m_a U^2 {\bf b} {\bf b} + \mu B_0 
( {\bf I} - {\bf b} {\bf b} ) ]
$.
    The expression of Eq.~(\ref{qNA}) coincides with 
that of the neoclassical radial heat flux in terms of the heat stress tensor.~\cite{H&S} 
    The turbulent heat flux in Eq.~(\ref{qA}) takes the same form as that 
given by the conventional gyrokinetic theory.~\cite{Sugama1996}

   The radial component of ${\bf Q}^{\rm C}$ in Eq.~(\ref{Q}) is ensemble-averaged to yield 
\begin{equation}
\label{QCs}
\langle \langle {\bf Q}^{\rm C} \cdot \nabla s \rangle \rangle
=
(q^{\rm C}_a)^s + \frac{5}{2} T_{a0} (\Gamma^{\rm C}_a)^s
+ {\cal O}(\delta^3)
,
\end{equation}
   where the radial classical heat flux is given by 
\begin{eqnarray}
\label{qC}
(q_a^{\rm C})^s 
& = & 
\sum_b
\left\langle \frac{m_a c}{e_a B_0}
\int dU \int d\mu \int d\xi \; D_a
[({\bf v} \times {\bf b} )\cdot \nabla s]
\right. 
\nonumber \\ & & 
\mbox{}
\times 
[
C_{ab}^p (\widetilde{f}_{a1}, F_{bM})
+ 
C_{ab}^p (F_{aM}, \widetilde{f}_{b1})
]
\nonumber \\ & & 
\left.
\mbox{}
\times
\left(
\frac{1}{2} m_a U^2 + \mu B_0
- \frac{5}{2} T_{a0}
\right)
\right\rangle 
\nonumber \\ 
& = & 
T_{a0}
\left\langle \frac{c}{e_a B_0}
( {\bf F}_{a2} \times {\bf b} )
\cdot \nabla s
\right\rangle 
.
\end{eqnarray}
%
   Here, ${\bf F}_{a2} \equiv \int d^3 {\bf v} 
(m_a v^2/2T_a - 5/2)  m_a {\bf v} C_a^p$ is the 
collisional heat friction. 
    The expression of the classical heat flux $(q_a^{\rm C})^s$ in Eq.~(\ref{qC}) 
agrees with the conventional one~\cite{H&S} 
and it immediately gives the classical heat 
transport equation relating $(q_a^{\rm C})^s$ to the gradient forces in the same way 
as mentioned after Eq.~(\ref{GammaC2}) for 
the classical particle flux $(\Gamma_a^{\rm C})^s$. 

  Now, Eq.~(\ref{Econs2a}) is rewritten as 
\begin{eqnarray}
\label{Econs3}
& & 
\frac{\partial }{\partial t}
\left (V' \left[
\frac{3}{2}\sum_a n_{a0} T_{a0}
+ \frac{B_0^2}{8\pi}
\right] \right)
\nonumber \\ 
& & 
+ \frac{\partial}{\partial s}
\left( V' \left[ \sum_a 
\left( 
(q_a)^s + 
\frac{5}{2} T_{a0} (\Gamma_a)^s
\right)
+ 
\left\langle 
{\bf S}^{\rm (Poynting)} \cdot \nabla s 
\right\rangle
\right. \right.
\nonumber \\ & & 
 \left. \left. \mbox{}
 - 
 \left(
\frac{3}{2}\sum_a n_{a0} T_{a0}
+ \frac{B_0^2}{8\pi}
\right)
\langle {\bf u}_s 
 \cdot \nabla s \rangle 
 \right] \right)
\nonumber \\ & & 
=
V' 
\sum_a
\left\langle 
\int d U \int d\mu \int d\xi \; D_a 
{\cal S}_a 
\left( \frac{1}{2} m_a U^2 + \mu B_0 \right)
\right\rangle
\nonumber \\ & &
\hspace{5mm} \mbox{}
+ {\cal O}(\delta^3)
,
\end{eqnarray}
   where the total radial heat flux is given by the sum of the 
classical, neoclassical, and 
turbulent parts as 
\begin{eqnarray}
(q_a)^s 
& = & 
(q_a^{\rm (gc)})^s +  (q_a^{\rm C})^s
\nonumber \\
& = &
(q_a^{\rm NA})^s  + (q_a^{\rm A})^s 
+ (q_a^{\rm C})^s 
,
\end{eqnarray}
   and ${\bf S}^{\rm (Poynting)} \equiv (c/4\pi) \langle {\bf E} \rangle_{\rm ens} 
\times {\bf B}_0$ represents the nonturbulent part of the Poynting vector. 
    Using the relation 
$
\langle
\partial 
( B_0^2 / 8\pi ) /\partial t
\rangle 
=
- (V')^{-1} \partial  ( V' \langle {\bf S}^{\rm (Poynting)} \cdot \nabla s \rangle )
/\partial s
-  
\langle
{\bf J}_0 \cdot \langle {\bf E} \rangle_{\rm ens}
\rangle 
$
shown in Ref.~18,
we also obtain 
\begin{eqnarray}
\label{Econs4}
& & 
\frac{\partial }{\partial t}
\left (V' 
\frac{3}{2}\sum_a n_{a0} T_{a0}
\right)
+  \frac{\partial}{\partial s}
\left( V' \left[ \sum_a 
\left( 
(q_a)^s + 
\frac{5}{2} T_{a0} (\Gamma_a)^s
\right)
\right. \right.
\nonumber \\ & & 
\mbox{}
\nonumber \\ & & 
 \left. \left. \mbox{}
 - 
\frac{3}{2}\sum_a n_{a0} T_{a0} \; 
\langle {\bf u}_s 
 \cdot \nabla s \rangle 
 \right] \right)
\nonumber \\ & & 
=  V' \langle {\bf J}_0 \cdot
\langle {\bf E} \rangle_{\rm ens} \rangle
+
V' 
\sum_a
\left\langle 
\int d U \int d\mu \int d\xi \; D_a 
{\cal S}_a 
\right.
\nonumber \\ & & 
\hspace*{3mm}
\mbox{} \left. 
\times
\left( \frac{1}{2} m_a U^2 + \mu B_0 \right)
\right\rangle
+ {\cal O}(\delta^3)
. 
\end{eqnarray}
    Equations~(\ref{Econs3}) and (\ref{Econs4}) take the well-known forms of the energy balance equations~\cite{Helander} except that the terms associated with the electric field energy and the kinetic energies due to the fluid velocities are neglected here as small quantities of higher order in $\delta$.

\subsection{Ensemble-averaged toroidal angular momentum balance equation}

   The ensemble-averaged toroidal angular momentum balance equation is 
written as 
\begin{eqnarray}
\label{tamc2}
& & 
\frac{\partial }{\partial t}
\left( V' 
\left\langle \left[ \sum_a n_{a0} m_a  ( u_{a\parallel} {\bf b} + {\bf u}_E ) 
+ \frac{{\bf S}^{\rm (Poynting)}}{c^2} \right] \cdot {\bf e}_\zeta \right\rangle
\right)
\nonumber \\ & & 
+  \frac{\partial}{\partial s}
\left( V' \left[ \sum_a \left\{  
(\Pi_a^{\rm NA})^s + (\Pi_a^{\rm A})^s + (\Pi_a^{\rm C})^s
\right. \right. \right.
\nonumber \\ & & 
 \left. \mbox{} 
\hspace*{-5mm}
 - 
\left\langle \left[ \sum_a n_{a0} m_a  ( u_{a\parallel} {\bf b} + {\bf u}_E ) 
+ \frac{{\bf S}^{\rm (Poynting)}}{c^2} \right] \cdot {\bf e}_\zeta
( {\bf u}_s \cdot \nabla s ) \right\rangle \right\}
\nonumber \\ & & 
\left. \left. \mbox{}
- \frac{1}{4\pi} \left\langle \left\langle \nabla s \cdot 
\left[
\hat{\bf E}_L \hat{\bf E}_L + \hat{\bf B} \hat{\bf B} 
+ ( \nabla \times \hat{\bf B} ) \hat{\bf A}
\right] \cdot {\bf e}_\zeta
\right\rangle \right\rangle
\right] \right)
\nonumber \\ & & 
=  
\sum_a
\left\langle 
\int d U \int d\mu \int d\xi \; D_a 
{\cal S}_a  
m_a U b_\zeta 
\right\rangle
+ {\cal O}(\delta^3)
, 
\end{eqnarray}
   where $u_{a\parallel}$ represents the nonturbulent part of the parallel fluid velocity for particle species $a$ defined by $n_{a0} u_{a\parallel} \equiv 
\int dU \int d\mu \int d\xi \langle F_{a1} \rangle_{\rm ens} U$ and 
${\bf u}_E \equiv c \langle {\bf E} \rangle_{\rm ens}
\times {\bf b}/B_0$ is the nonturbulent part of 
the ${\bf E}\times {\bf B}$ drift velocity. 
   Equation~(\ref{tamc2}) is derived from 
Eq.~(\ref{tamc1a}) following the same procedures as shown in 
Ref.~18 except that 
the additional transport flux $\Pi_a^{\rm C}$ 
defined in Eq.~(\ref{PiaC}) 
and 
the external momentum source  
are newly included in the present case. 

   On the left-hand side of Eq.~(\ref{tamc2}), 
 the terms including $( u_{a\parallel} {\bf b} + {\bf u}_E )$ and 
${\bf S}^{\rm (Poynting)}$ are of ${\cal O}(\delta^3)$ although they are written down 
to explicitly show the inertia-term part. 
   The nonturbulent and turbulence-driven parts of the radial flux of 
the toroidal angular momentum are defined by 
\begin{eqnarray}
\label{PiNA}
(\Pi_a^{\rm NA})^s & =  &
\left\langle \int dU \int d\mu \int d\xi \;
D_a \langle F_{a1} \rangle_{\rm ens}
\right. 
\nonumber \\ & & \mbox{}
\left. 
\times 
m_a U b_\zeta 
\langle 
{\bf v}_a^{\rm (gc)} \rangle_{\rm ens}
\cdot \nabla s
\right\rangle 
,
\end{eqnarray}
   and 
\begin{eqnarray}
\label{PiA}
(\Pi_a^{\rm A})^s 
& = &
\left\langle \left\langle 
\int dU \int d\mu \int d\xi \;
D_a \hat{h}_a
\right. \right.
\nonumber \\ & & 
\left. \left. 
\mbox{} \times
m_a  ( 
U {\bf b}
+ {\bf v}_{a0\perp}
) 
\cdot {\bf e}_\zeta
( \hat{\bf v}_a^{\rm (gc)} \cdot \nabla s )
\right\rangle \right\rangle
, 
\hspace*{5mm}
\end{eqnarray}
   respectively. 
   It is shown in Appendix~D that 
Eq.~(\ref{PiC}) 
is ensemble-averaged to give 
\begin{equation}
\label{PiC_av}
\left\langle  \left\langle
(\Pi^{\rm C})^s 
\right\rangle \right\rangle
=
\sum_a \Pi_a^{\rm C} 
+ {\cal O} (\delta^3)
,
\end{equation}
    where the radial transport flux of the toroidal angular momentum for species $a$ 
due to the collision term and finite gyroradii is defined by 
\begin{eqnarray}
\label{PiaC}
& & 
 \Pi_a^{\rm C} = 
-
\sum_b
\left\langle 
\frac{m_a c  |\nabla s|^2}{2 e_a B_0}
\frac{\partial \chi}{\partial s}
\int dU \int d\mu \int d\xi \; D_a \mu
\right. 
\nonumber  \\
& & 
\left. \hspace*{10mm}
\times 
\left[
C_{ab}^p (\langle F_{a1} \rangle_{\rm ens}, F_{bM} )
+ C_{ab}^p ( F_{aM}, \langle F_{b1} \rangle_{\rm ens}  )
\right]
\right\rangle 
. 
\nonumber \\ & & 
\end{eqnarray}
The expressions for the toroidal momentum fluxes 
shown in Eqs.~(\ref{PiNA})--(\ref{PiaC}) agree with 
those given by conventional recursive formulations in 
Refs.~33--35. 
[Since the so-called high-flow ordering is used in 
Refs.~33 and 34, 
the expressions for the toroidal momentum fluxes in it reduce to those 
in the present work in the low-flow-speed limit.]
As argued in Refs.~18 and 35, 
when there exists the up-down symmetry of the background 
magnetic field, all toroidal momentum fluxes vanish to ${\cal O}(\delta^2)$ 
and the nontrivial toroidal momentum balance equation is of ${\cal O}(\delta^3)$. 
   In this case, 
gyrokinetic systems equations of higher-order accuracy in $\delta$ 
are required for the correct derivation of this ${\cal O}(\delta^3)$ 
 toroidal momentum balance equation to determine the profile of the 
radial electric field~\cite{Calvo} 
although we should note, at the same time, that 
the radial electric field is not necessary to determine 
the particle and energy transport fluxes to the lowest order 
in $\delta$.~\cite{Sugama2011}

\section{CONCLUSIONS}

In this paper, particle, energy, and toroidal momentum balance equations 
including collisional and turbulent transport fluxes are systematically derived 
from the gyrokinetic Boltzmann-Poisson-Amp\`{e}re  system of equations. 
   Considering an imaginary collisionless system, for which the distribution functions and electromagnetic fields coincide instantaneously with those for the considered collisional system, 
and expressing the variation of 
the action integral for the collisionless system in terms of the solution to 
the governing equations for the collisional system 
clarify effects of the collision and external source terms on 
the collisionless conservation laws derived from Noether's theorem. 
   The gyrokinetic collision operator is newly presented, by which  
the collisional changes in the velocity-space integrals of the gyrocenter Hamiltonian 
and the canonical toroidal angular momentum can be written  
in the conservative (or divergence) forms. 
   It is confirmed that, to the lowest order in the 
normalized gyroradius, 
the ensemble-averaged fluxes in the derived particle, energy, and toroidal 
angular momentum balance equations can be written by the sum of  
conventional expressions of classical, neoclassical, and turbulent 
transport fluxes.
  The extension of the present work to the case of the high-flow ordering 
remains as a future task.

\begin{acknowledgments}
This work was supported in part by NIFS/NINS under 
the Project of
Formation of International Network for 
Scientific Collaborations, 
the NIFS Collaborative Research Programs 
(NIFS14KNTT026, NIFS15KNTT031), 
and in part by the Japanese Ministry
of Education, Culture, Sports, Science and Technology
(Grant No.\ 26820398).   
\end{acknowledgments}

\appendix

\section{COORDINATE TRANSFORMATION}

We  consider the transformation of the phase-space coordinates in this Appendix, 
where the subscript representing the particle species is omitted as far as it is unnecessary. 
In terms of the position ${\bf x}$ and the velocity ${\bf v}$ of a given particle, 
we define the parallel velocity $v_\parallel  = {\bf v} \cdot {\bf b}({\bf x}, t)$,  
the perpendicular velocity ${\bf v}_\perp = {\bf v} - v_\parallel {\bf b}$, and 
the zeroth-order magnetic moment, 
\begin{equation}
\mu_0 = \frac{m v_\perp^2}{2 B_0({\bf x}, t)}
\end{equation}
   where the equilibrium field at position ${\bf x}$ and time $t$ 
is denoted by ${\bf B}_0 ({\bf x}, t) = B_0 ({\bf x}, t) {\bf b}({\bf x}, t)$.   
  We also define the zeroth-order gyrophase  by 
$\xi_0 = \tan^{-1} [({\bf v}\cdot {\bf e}_1)/({\bf v}\cdot {\bf e}_2)$ 
where  $({\bf e}_1, {\bf e}_2, {\bf b})$ are unit vectors 
which form a right-handed orthogonal system at $({\bf x}, t)$. 
   Then, the gyrocenter coordinates ${\bf Z} = ({\bf X}, U, \mu, \xi)$ are 
represented in terms of the particle coordinates 
${\bf z} = ({\bf x}, v_\parallel, \mu_0, \xi_0)$ as 
\begin{equation}
\label{p2g}
({\bf X}, U, \mu, \xi) =  
({\bf x}, v_\parallel, \mu_0, \xi_0) + 
(\Delta {\bf x}, \Delta v_\parallel, \Delta  \mu_0, \Delta  \xi_0)
, 
\end{equation}
   where 
\begin{eqnarray}
\label{Deltaz}
\Delta {\bf x}
& = & 
 - \mbox{\boldmath$\rho$} 
 + {\cal O} (\delta^2)
\nonumber \\ 
\Delta v_\parallel
& = & 
- v_\parallel {\bf b} \cdot \nabla  {\bf b} \cdot 
\mbox{\boldmath$\rho$} 
- \frac{1}{4} ( 3 \mbox{\boldmath$\rho$}  \cdot \nabla  {\bf b} \cdot {\bf v}_\perp
- {\bf v}_\perp \cdot \nabla  {\bf b} \cdot \mbox{\boldmath$\rho$} )
\nonumber \\ 
&  & 
\mbox{} + \frac{e}{mc} A_{1\parallel} 
+  {\cal O} (\delta^2)
\nonumber \\ 
\Delta \mu_0
& = & 
\frac{m}{B_0} 
\left[ v_\parallel^2 {\bf b} \cdot \nabla {\bf b} \cdot 
\mbox{\boldmath$\rho$} + \frac{v_\parallel}{4} 
( 3 \mbox{\boldmath$\rho$}  \cdot \nabla  {\bf b} \cdot {\bf v}_\perp
- {\bf v}_\perp \cdot \nabla  {\bf b} \cdot \mbox{\boldmath$\rho$} )
\right. 
\nonumber  \\ & & 
\left. \mbox{} 
+ \frac{v_\perp^2}{2 B_0} \mbox{\boldmath$\rho$} \cdot \nabla B_0
\right]
+ \frac{e}{cB_0}{\bf v}_\perp \cdot {\bf A}_{1\perp}
+ \frac{e}{B_0} \widetilde{\psi} 
\nonumber \\ & & \mbox{}
+  {\cal O} (\delta^2)
\nonumber \\ 
\Delta \xi_0
& = & 
 \frac{1}{\Omega B_0} ( {\bf  v}_\perp \cdot \nabla B_0 )
- \frac{\Omega v_\parallel}{4 v_\perp^2} \mbox{\boldmath$\rho$}
 \cdot \nabla {\bf b} \cdot \mbox{\boldmath$\rho$} 
\nonumber \\ 
& & \mbox{}
+ \frac{v_\parallel}{4 \Omega v_\perp^2} 
{\bf v}_\perp \cdot \nabla {\bf b} \cdot {\bf v}_\perp 
+ \frac{v_\parallel^2}{\Omega v_\perp^2} 
( {\bf b} \cdot \nabla {\bf b} \cdot {\bf v}_\perp )
\nonumber \\ 
& & \mbox{}
- \frac{\Omega}{v_\perp^2} \mbox{\boldmath$\rho$}
 \cdot \nabla {\bf  v}_\perp \cdot \mbox{\boldmath$\rho$} 
+ \frac{e}{c m v_\perp^2} {\bf b} 
\cdot ( {\bf A}_1 \times {\bf v} )
\nonumber \\ 
& & \mbox{}
 - \frac{e}{B_0} 
\left( \int 
\frac{\partial \widetilde{\psi}}{\partial \mu_0} d\xi_0 \right)
+  {\cal O} (\delta^2)
. 
\end{eqnarray}
   The formulas for $\Delta v_\parallel$ 
$\Delta \mu_0$, and $\Delta \xi_0$ in Eq.~(\ref{Deltaz}) are 
obtained by 
combining the guiding center and gyrocenter coordinate 
transformations.~\cite{B&H,Sugama2000,Littlejohn1981} 
   Here, effects of the background electric field and turbulent 
electromagnetic fields 
are included through $\widetilde{\psi}$ [see Eq.~(\ref{psia})] and 
${\bf A}_1$. 
   When the background electric field and turbulent 
electromagnetic fields vanish, 
$\Delta v_\parallel$ 
$\Delta \mu_0$, and $\Delta \xi_0$ in Eq.~(\ref{Deltaz}) 
agree with the results in Ref.~37.

Denoting the coordinate transformation by ${\cal T}$, 
Eq.~(\ref{p2g}) is rewritten as 
\begin{equation}
\label{Z=Tz}
{\bf Z} = {\cal T} ({\bf z})
= {\bf z} + \Delta {\bf z}
.
\end{equation}
   An arbitrary scalar field ${\cal A}$ on the phase space can be expressed 
in terms of either the gyrocenter coordinates 
${\bf Z} = ( {\bf X}, U, \mu, \xi )$ 
or the particle coordinates 
${\bf z} = ({\bf x}, v_\parallel, \mu_0, \xi_0)$ 
as 
\begin{equation}
\label{AgAp}
{\cal A}^g ({\bf Z}) =  {\cal A}^p ({\bf z}) 
. 
\end{equation}
   Using Eqs.~(\ref{Z=Tz}), (\ref{AgAp}), and the Taylor series expansion, 
we obtain
\begin{eqnarray}
\label{Ap_series}
& & 
{\cal A}^p ({\bf z}) = ({\cal T}^*  {\cal A}^g)({\bf z})  
\equiv  {\cal A}^g ({\cal T}({\bf z})) = {\cal A}^g ({\bf z}+ \Delta {\bf z}) 
\nonumber \\ 
&  & 
=
\sum_{n=0}^\infty \frac{1}{n!}
\sum_{i_1,\cdots,i_n}
\Delta z^{i_1} \cdots \Delta z^{i_n}
\frac{\partial^n {\cal A}^g ( {\bf z} )}{
\partial z^{i_1} \cdots \partial z^{i_n}}
, 
\end{eqnarray}
   where ${\cal T}^*{\cal A}^g$ denotes the pullback 
transformation of ${\cal A}^g$ by ${\cal T}$. 
   Using the inverse transformation ${\cal T}^{-1}$, we also have 
${\cal A}^g ({\bf Z}) = ({\cal T}^{-1*}{\cal A}^p)({\bf Z})
\equiv   {\cal A}^p ({\cal T}^{-1} ({\bf Z})) $.

   The Jacobians $D^p$ and $D^g$ for the two coordinate 
systems ${\bf z}$ and ${\bf Z}$ are related to each other 
by
\begin{equation}
D^p ({\bf z}) =  
\det 
\left[ 
\frac{\partial ({\bf Z})}{\partial ({\bf z})}
\right]
D^g ({\bf Z}) 
,
\end{equation}
where $\partial ({\bf Z})/\partial ({\bf z})$ denotes the Jacobian matrix. 
   Then, we use the following formula, 
\begin{equation}
\label{delta_z}
\delta^6 ( {\bf z} + \Delta {\bf z}
- {\bf Z} )
= \sum_{n = 0}^\infty
\frac{1}{n!} \sum_{i_1, \cdots, i_n}
\Delta z_{i_1} \cdots \Delta z_{i_n} 
\frac{\partial^n \delta^6 ({\bf z} - {\bf Z} )}{\partial z_{i_1} \cdots \partial z_{i_n} }
, 
\end{equation}
   and partial integrals to derive the relation between the expressions of the scalar density 
${\cal D} {\cal A}$ in the gyrocenter and  particle coordinate systems as 
\begin{eqnarray}
\label{DgAg}
& & 
D^g ({\bf Z}) {\cal A}^g ({\bf Z}) 
\nonumber \\ 
& = & 
\int d^6 {\bf Z}' \;
\delta^6 ( {\bf Z}' - {\bf Z} )D^g ({\bf Z}')  {\cal A}^g ({\bf Z}') 
\nonumber \\ 
& = & 
\int d^6 {\bf z} \;
\delta^6 ( {\bf z} + \Delta {\bf z} - {\bf Z} ) D^p ({\bf z})  {\cal A}^p ({\bf z}) 
\nonumber \\ 
& = & 
\sum_{n=0}^\infty \frac{(-1)^n}{n!}
\sum_{i_1,\cdots,i_n}
\left[
\frac{\partial^n 
\left[
\Delta z^{i_1} \cdots \Delta z^{i_n}
D^p ({\bf z}) {\cal A}^p ( {\bf z} )
\right]
}{
\partial z^{i_1} \cdots \partial z^{i_n}}
\right]_{{\bf z} = {\bf Z}}
, 
\nonumber \\ & & 
\end{eqnarray}
   where the replacement of ${\bf z}$ with ${\bf Z}$ is 
represented by 
$[ \cdots ]_{{\bf z} = {\bf Z}} 
\equiv  \int d^6 {\bf z} \;
\delta^6 ( {\bf z} - {\bf Z} ) \cdots $.

\section{COLLISION OPERATOR IN GYROCENTER COORDINATES}

   We can regard the collision term as a scalar field $C$ on the phase space. 
When using the particle coordinates, 
we represent the collision term for collisions between species $a$ and $b$
by $C_{ab}^p$.  
A well-established collision operator $C_{ab}^p(f_a, f_b)$ for the 
particle distribution functions $f_a$ and $f_b$ is known as 
the Landau operator [see, for example, Eq.~(3.22) in Ref.~32]. 
   Then, the collision term $C_{ab}^g$ represented in the gyrocenter coordinates 
is related to $C_{ab}^p$ by 
\begin{equation}
\label{scalar_C}
C_{ab}^g (F_a, F_b) = {\cal T}_a^{-1*} 
C_{ab}^p ({\cal T}_a^* F_a, {\cal T}_b^* F_b )
,
\end{equation}
   where the distribution function for species $a$ $(b)$  in the particle coordinates 
is written as the pullback $f_a = {\cal T}_a^* F_a$ $(f_b = {\cal T}_b^*F_b)$ of 
that in the gyrocenter coordinates $F_a$ $(F_b)$ by 
the coordinate transformation ${\cal T}_a$ $({\cal T}_b)$ described in Appendix~A, 
and ${\cal T}_a^{-1*}$ transforms 
the collision term as a function of the particle coordinates 
into that of the gyrocenter coordinates. 

    In order to see collisional effects on conservation laws, it is convenient to 
represent the collision term in the gyrocenter coordinate using 
the transformation formula for the scalar density $D_a C_{ab}$ rather than 
that for the scalar $C_{ab}$ shown in Eq.~(\ref{scalar_C}). 
   Using Eq.~(\ref{DgAg}), we can derive  
\begin{eqnarray}
\label{DCgA}
& & 
D_a^g ({\bf Z}_a)
C_{ab}^g [F_a, F_b]({\bf Z}_a) 
{\cal A}_a^g ({\bf Z}_a) 
\nonumber \\ 
& = & 
\sum_{n=0}^\infty \frac{(-1)^n}{n!}
\sum_{i_1,\cdots,i_n}
\nonumber \\ 
&  & 
\hspace*{-5mm}
\times 
\left[
\frac{\partial^n 
\left[
\Delta z_a^{i_1} \cdots \Delta z_a^{i_n}
D_a^p ({\bf z}_a)
C_{ab}^p [f_a, f_b]({\bf z}_a) {\cal A}_a^p ({\bf z}_a) 
\right]
}{
\partial z_a^{i_1} \cdots \partial z_a^{i_n}}
\right]_{{\bf z}_a = {\bf Z}_a}
,
\nonumber \\ & & 
\end{eqnarray}
   where ${\cal A}_a$ is an arbitrary scalar field depending on  
particle species and 
$f_a = {\cal T}_a^* F_a$ is rewritten by using  Eq.~(\ref{Ap_series}) as 
\begin{equation}
\label{fzFTz}
f_a ({\bf z}_a)  
= 
\sum_{n=0}^\infty \frac{1}{n!}
\sum_{i_1,\cdots,i_n}
\Delta z_a^{i_1} \cdots \Delta z_a^{i_n}
\frac{\partial^n F_a ( {\bf z}_a )}{
\partial z_a^{i_1} \cdots \partial z_a^{i_n}}
.
\end{equation}
   Then, the gyrocenter representation of the collision operator $C_{ab}^g$ 
acting on $F_a$ and $F_b$ is obtained by Eq.~(\ref{DCgA})  
with putting ${\cal A}^g = {\cal A}^p = 1$ 
and using Eq.~(\ref{fzFTz}) to 
express $f_a$ and $f_b$ in terms of  $F_a$ and $F_b$, respectively.  
   Integrating Eq.~(\ref{DCgA}) with respect to $(U, \mu, \xi)$ 
and taking the summation over species $b$ 
yield 
\begin{eqnarray}
\label{intvDCgA}
& & 
\int dU \int d\mu \int d\xi \;  
D_a^g ({\bf Z})
C_a^g ({\bf Z}) 
{\cal A}_a^g ({\bf Z}) 
\nonumber \\ 
&  & 
= 
\left[
\int d^3 {\bf v} \;  
C_a^p ({\bf z}) {\cal A}_a^p ({\bf z}) 
\right]_{{\bf z} = {\bf Z}}
- \nabla \cdot {\bf J}_{Aa}^{\rm C},
\nonumber \\ 
&  & 
\end{eqnarray}
   where $C_a^g = \sum_b C_{ab}^g$ and  
$\nabla = \partial/\partial {\bf X}$ are used  
and $\int d^3 {\bf v} = 
\int dv_\parallel \int d\mu_0 \int d\xi_0 \: D_a^p ({\bf z})$ 
denotes the velocity-space integral using the particle coordinates. 
   Here, the transport flux ${\bf J}_{Aa}^{\rm C}$  
of the quantity ${\cal A}_a$ due 
to collisions and finite gyroradii of particles
is defined by
\begin{eqnarray}
\label{JAC}
& & 
{\bf J}_{Aa}^{\rm C}
({\bf X})
\nonumber \\ 
& = & 
\sum_{n=0}^\infty 
\frac{(-1)^n}{(n+1)!}
\sum_{i_1,\cdots,i_n}
\frac{\partial^n}{
\partial X^{i_1} \cdots \partial X^{i_n}}
\nonumber \\ 
&  & 
\times
\left[
\int d^3 {\bf v} \; \Delta {\bf x}_a
\Delta x_a^{i_1} \cdots \Delta x_a^{i_n}
C_a^p ({\bf z})
{\cal A}_a^p ({\bf z})
\right]_{{\bf x} = {\bf X}}
\nonumber \\ 
& = & 
\left[
\int d^3 {\bf v} \;
\Delta {\bf x}_a
C_a^p ({\bf z})
{\cal A}_a^p ({\bf z})
\right]_{{\bf x} = {\bf X}}
+ \cdots
.
\end{eqnarray}
The integral of an arbitrary scalar field ${\cal A}_a$ over the whole phase space 
is written in either the gyrocenter or particle coordinate system as 
\begin{equation}
\label{intDCA}
\int d^6 {\bf Z} \; 
D^g_a ({\bf Z}) C^g_a ({\bf Z}) {\cal A}^g_a ({\bf Z}) 
=
\int d^6 {\bf z} \; 
D^p_a ({\bf z}) C^p_a ({\bf z}) {\cal A}^p_a ({\bf z}) 
. 
\end{equation}

   For the case of ${\cal A}_a = 1$,  
Eqs.~(\ref{intvDCgA}) and (\ref{JAC}) reduce to 
\begin{equation} 
\label{intvDCg}
\int dU \int d\mu \int d\xi \; D_a^g ({\bf Z}) C_a^g 
({\bf Z})
=
- \nabla \cdot \mbox{\boldmath$\Gamma$}_a^{\rm C} ({\bf X}),
\end{equation}
   and 
\begin{eqnarray}
\label{GammaC}
& & 
\mbox{\boldmath$\Gamma$}_a^{\rm C} 
({\bf X})
\nonumber \\ 
& = & 
\sum_{n=0}^\infty 
\frac{(-1)^n}{(n+1)!}
\sum_{i_1,\cdots,i_n}
\frac{\partial^n}{
\partial X^{i_1} \cdots \partial X^{i_n}}
\nonumber \\ 
&  & 
 \times
\left[
\int d^3 {\bf v} \; \Delta {\bf x}_a
\Delta x_a^{i_1} \cdots \Delta x_a^{i_n}
C_a^p ({\bf z})
\right]_{{\bf z} = {\bf Z}}
\nonumber \\ 
& = & 
\left[ 
\int d^3 {\bf v} \;
  \Delta {\bf x}_a
C_a^p ({\bf z})
\right]_{{\bf x} = {\bf X}}
+ \cdots
, 
\end{eqnarray}
respectively, where 
$\int d^3 {\bf v} \: C_a^p({\bf z}) = 0$ is used. 
   Here, $\mbox{\boldmath$\Gamma$}_a^{\rm C}$ is regarded as the 
classical particle flux which occurs due to collisions and finite gyroradii. 
   In fact, using $\Delta {\bf x}_a \simeq -\mbox{\boldmath$\rho$}_a$, we see that the 
primary term of $\mbox{\boldmath$\Gamma$}_a^{\rm C}$ shown in  
the last line of Eq.~(\ref{GammaC}) is identical to 
the conventional definition of the classical particle flux 
$\mbox{\boldmath$\Gamma$}_a^{\rm cl} 
\equiv (c/e_a B_0) {\bf F}_{a1} \times {\bf b}$, 
where ${\bf F}_{a1} \equiv \int d^3 {\bf v} \: m_a {\bf v} \:  C_a^p$ is the 
collisional friction force. 
Thus, we have 
$\mbox{\boldmath$\Gamma$}_a^{\rm C} = 
\mbox{\boldmath$\Gamma$}_a^{\rm cl} [1 + {\cal O}(\delta)]$.

Let us take the kinetic energy of the particle as ${\cal A}_a$ and 
put ${\cal A}_a^p = \frac{1}{2} m_a v_a^2
= \frac{1}{2} m_a v_{\parallel a}^2 + \mu_{0a} B_0 ({\bf x}_a)$.  
    Then, it is written  in terms of  the gyrocenter coordinates as 
\begin{eqnarray}
\label{kin-en}
 & & 
{\cal A}_a^g 
=
{\cal T}_a^{-1*} \left( \frac{1}{2} m_a v_a^2 \right)
\nonumber \\ 
& = & 
\frac{1}{2} m_a U_a^2 + \mu_a  B_0 ({\bf X}_a)
+ \mu_a \mbox{\boldmath$\rho$}_a \cdot \nabla B_0 ({\bf X}_a)
\nonumber \\ 
&  & \mbox{}
- m_a U_a (\Delta v_{\parallel a})_{{\bf z}_a = {\bf Z}_a}
-  (\Delta \mu_{0a})_{{\bf z}_a = {\bf Z}_a} B_0 ({\bf X}_a)
+ \cdots
\nonumber \\ 
&  = & 
\frac{1}{2} m_a U_a^2 + \mu_a  B_0 ({\bf X}_a) 
+ e_a \langle \psi_a ({\bf Z}_a) \rangle_{\xi_a}
  - e_a   ( {\cal T}_a^{-1*} \phi ) ({\bf Z}_a)
\nonumber \\ 
& & \mbox{}
+ {\cal O}(\delta^2)
, 
\end{eqnarray}
where the inverse ${\cal T}_a^{-1}$ of the transformation ${\cal T}_a$ given 
by Eq.~(\ref{Deltaz}) is used. 
     In this case, taking the summation of Eq.~(\ref{intvDCgA}) over species $a$ 
and using the conservation property 
$\sum_a \int d^3 {\bf v} \;  C_a^p \frac{1}{2} m_a v^2 = 0$, 
we have 
\begin{eqnarray}
\label{intvDCgKP}
& & 
\sum_a
\int dU \int d\mu \int d\xi \;  
D_a^g ({\bf Z})
C_a^g ({\bf Z}) 
T_a^{-1*} \left( \frac{1}{2} m_a v^2 \right)
\nonumber \\ 
&  & 
= 
- \nabla \cdot {\bf Q}^{\rm C},
\end{eqnarray}
   where ${\bf Q}^{\rm C}$ represents 
the transport flux of the total kinetic energy due to collisions and 
finite gyroradii defined by 
\begin{eqnarray}
\label{JKE}
& & 
{\bf Q}^{\rm C}
({\bf X})
= \sum_a
\sum_{n=0}^\infty 
\frac{(-1)^n}{(n+1)!}
\sum_{i_1,\cdots,i_n}
\frac{\partial^n}{
\partial X^{i_1} \cdots \partial X^{i_n}}
\nonumber \\ 
&  & 
\hspace*{5mm}
\times
\left[
\int d^3 {\bf v} \; \Delta {\bf x}_a
\Delta x_a^{i_1} \cdots \Delta x_a^{i_n}
C_a^p ({\bf z})
\frac{1}{2} m_a v^2
\right]_{{\bf x} = {\bf X}}
\nonumber \\ 
&  & 
=
\sum_a
\left[
\int d^3 {\bf v} \;
\Delta {\bf x}_a
C_a^p ({\bf z})
\frac{1}{2} m_a v^2
\right]_{{\bf x} = {\bf X}}
+ \cdots
.
\end{eqnarray}
    To the lowest order in $\delta$, 
the collisional energy flux ${\bf Q}^{\rm C}$ 
is approximately written as  ${\bf Q}^{\rm C} \simeq 
\sum_a ({\bf q}_a^{\rm cl} + \frac{5}{2}T_a 
\mbox{\boldmath$\Gamma$}_a^{\rm cl})$.  
    Here, the classical heat flux for species $a$ is defined by  
${\bf q}_a^{\rm cl} \equiv
(c T_a /e_a B_0) {\bf F}_{a2} \times {\bf b}$, 
where ${\bf F}_{a2} \equiv \int d^3 {\bf v} 
(m_a v^2/2T_a - 5/2)  m_a {\bf v} C_a^p$ is the 
collisional heat friction. 
   We note from Eq.~(\ref{kin-en}) that the expression of the kinetic energy 
in the gyrocenter coordinates should be generally given by the infinite 
series expansion in $\delta$ 
in order for the gyrocenter velocity-space integral of the collisional rate of change in 
the kinetic energy to take the form of the divergence of the energy flux without 
any local source or sink terms. 
   In fact,  this energy conservation property is broken if we keep only the lowest order terms 
in Eq.~(\ref{kin-en}) and evaluate the gyrocenter velocity-space integral 
$
\sum_a
\int dU \int d\mu \int d\xi \;  
D_a^g ({\bf Z})
C_a^g ({\bf Z}) 
\left( \frac{1}{2} m_a U^2 + \mu B_0 ({\bf X}) \right)
$. 

The above-mentioned subtle relation between expressions of 
the collisional energy conservation properties in the particle and 
gyrocenter coordinate systems is also found when considering 
the collisional momentum conservation. 
   It should be recalled 
that the perturbative expansions in $\delta$ are truncated 
up to finite orders in deriving gyrokinetic equations as shown in Sec.~III 
although the conservative form of equations for the energy and the toroidal 
angular momentum are obtained even from these approximate equations
for the collisionless case since they are constructed based on the variational principle. 
    Thus, from the viewpoint of practical applications, it is desirable for 
the approximate collision operator in the gyrocenter coordinates 
to keep the conservation properties. 
    More rigorously speaking, we want the gyrokinetic collisional 
velocity-space integrals 
$
\sum_a
\int dU \int d\mu \int d\xi \;  
D_a^g ({\bf Z})
C_a^g ({\bf Z}) 
H_a ({\bf Z})
$
and 
$
\sum_a
\int dU \int d\mu \int d\xi \;  
D_a^g ({\bf Z})
C_a^g ({\bf Z}) 
(p_\zeta^c)_a^g({\bf Z})
$
to take the divergence forms and include no local source or sink terms  
where $H_a ({\bf Z})$ and 
$(p_\zeta^c)_a^g({\bf Z}) \equiv (e_a/c) A_{a\zeta}^*({\bf Z})$ are 
the gyrocenter Hamiltonian and the canonical toroidal angular momentum 
defined by 
Eqs.~(\ref{Ha}) and (\ref{pcz}), respectively. 
Here, it should be noted that 
not only kinetic parts of energy and toroidal momentum 
but also contributions from scalar and vector potentials are included 
in $H_a ({\bf Z})$ and $(e_a/c) A_{a\zeta}^*({\bf Z})$. 
   In Appendix~C, we find how to construct the approximate 
gyrokinetic collision operator, by which the two integrals mentioned above 
are written in the divergence forms. 

    We now consider  the entropy per unit volume defined in terms of 
the gyrocenter distribution functions as 
$S^g \equiv - \sum_a \int dU \int d\mu \int d\xi 
D_a^g ({\bf Z}) \log F_a ({\bf Z})$, in which 
the rate of change is given by $d S^g / d t = 
- \sum_a \int dU \int d\mu \int d\xi \:
D_a^g ({\bf Z}) \: [ \log F_a ({\bf Z})+1 ] \: (d F_a / d t )$. 
   Then,  the rate of change in $S^g$ due to collisions is 
obtained by putting ${\cal A}_a^g = -  [ \log F_a ({\bf Z})+1]$ 
in Eq.~(\ref{intvDCgA}) 
and taking the summation over species $a$ as
\begin{eqnarray}
\label{intvDCglogF}
& & 
- \sum_a
\int dU \int d\mu \int d\xi \;  
D_a^g ({\bf Z})
C_a^g ({\bf Z}) [\log F_a ({\bf Z}) + 1]
\nonumber \\ 
&  & 
= 
- \sum_a \left[
\int d^3 {\bf v} \;
C_a^p ({\bf z})
\log f_a ({\bf z}) 
\right]_{{\bf x} = {\bf X}}
- \nabla \cdot {\bf J}_S^{\rm C}
, 
\hspace*{10mm}
\end{eqnarray}
where $\int d^3 {\bf v} \; C_a^p ({\bf z}) = 0$ is used 
although we should recall that 
$\int dU \int d\mu \int d\xi D_a^g ({\bf Z}) C_a^g ({\bf Z})$ 
does not vanish generally as seen from Eq.~(\ref{intvDCg}). 
   It is well-known that, when Landau's collision operator 
is used for $C_a^p$, 
the collisional entropy production rate given by 
the first term on the right-hand side of 
Eq.~(\ref{intvDCglogF}) is nonnegative. 
   This is Boltzmann's H-theorem which proves the second law of thermodynamics. 
    The collisional transport flux ${\bf J}_S^{\rm C}$ 
of the entropy in Eq.~(\ref{intvDCglogF}) is defined by 
\begin{eqnarray}
\label{JS}
& & 
{\bf J}_S^{\rm C} ({\bf X})
\nonumber \\ 
& = & 
\sum_a
\sum_{n=0}^\infty 
\frac{(-1)^{n+1}}{(n+1)!}
\sum_{i_1,\cdots,i_n}
\frac{\partial^n}{
\partial X^{i_1} \cdots \partial X^{i_n}}
\nonumber \\ 
&  & 
\times
\left[
\int d^3 {\bf v} \; \Delta {\bf x}_a
\Delta x_a^{i_1} \cdots \Delta x_a^{i_n}
C_a^p ({\bf z})
[ \log f_a ({\bf z}) + 1 ]
\right]_{{\bf x} = {\bf X}}
\nonumber \\ 
& = & 
- \sum_a
\left[
\int d^3 {\bf v} \;
\Delta {\bf x}_a
C_a^p ({\bf z})
[ \log f_a ({\bf z}) + 1 ]
\right]_{{\bf x} = {\bf X}}
+ \cdots
.
\nonumber \\ & & 
\end{eqnarray}
It is shown that, to the lowest order in $\delta$, 
the collisional entropy transport flux is written as 
${\bf J}_S^{\rm C} = \sum_a ( S_{a0} 
{\bf u}_a^{\rm cl} + {\bf q}_a^{\rm cl} / T_a )
$ 
where  
the lowest-order entropy density $S_{a0}$ 
for species $a$ is given in terms of  
the local Maxwellian distribution function $F_{aM}$ as 
$S_{a0} \equiv -  \int dU \int d\mu \int d\xi  \;  F_{aM} \log F_{aM}$, 
and ${\bf u}_a^{\rm cl}$ is defined by ${\bf u}_a^{\rm cl} \equiv 
\mbox{\boldmath$\Gamma$}_a^{\rm cl} / n_a$. 
     Here, we note again that the infinite series expansion in $\delta$ 
as given by  Eq.~(\ref{DCgA}) is used in deriving 
Eq.~(\ref{intvDCglogF}). 
     When the expansion is truncated to finite order, 
the collisional entropy production term is represented by
$- \sum_a 
\int d^3 {\bf v} \;C_a^p \log f_a$ 
plus residual error terms of higher order in $\delta$, 
and thus the H-theorem is only approximately satisfied.

\section{COLLISION OPERATOR RELEVANT FOR GYROKINETIC CONSERVATION LAWS}

In this Appendix, we consider an approximate gyrokinetic collision operator 
instead of the one given by Eq.~(\ref{scalar_C}) 
[or Eq.~(\ref{DCgA}) with ${\cal A}^g = {\cal A}^p = 1$] 
in order to 
get the gyrokinetic collisional 
velocity-space integrals of energy and canonical toroidal momentum 
to take desirable conservative (or divergence) forms. 
    The approximate collision operator is written in the 
gyrocenter coordinates as 
\begin{eqnarray}
\label{DCg2}
& & 
D_a^g ({\bf Z}_a)
\langle C_{ab}^g [F_a, F_b]({\bf Z}_a) \rangle_{\xi_a}
\nonumber \\ 
& = &
\left\langle 
\left[
\sum_{n=0}^\infty \frac{1}{n!}
\right. \right.
\nonumber \\ 
&  & \mbox{} \times
\sum_{i_1,\cdots,i_n}
\frac{\partial^n 
\left[
\rho_a^{i_1} \cdots \rho_a^{i_n}
D_a^p ({\bf z}_a)
C_{ab}^p [f_a, f_b]({\bf z}_a)
\right]
}{
\partial x_a^{i_1} \cdots \partial x_a^{i_n}}
\nonumber \\ & & 
\mbox{}
- \frac{\partial}{\partial {\bf x}_a} 
\cdot  
\left[
\Delta  {\bf x}_a^{(2)} 
D_a^p ({\bf z}_a) C_{ab}^p [f_a, f_b]({\bf z}_a)
\right]
\nonumber \\ & & 
\mbox{}
- \frac{\partial}{\partial v_{\parallel a}} 
\left[
\Delta  v_{\parallel a} D_a^p ({\bf z}_a) C_{ab}^p [f_a, f_b]({\bf z}_a)
\right]
\nonumber \\ & & 
\left. \left.
\mbox{}
- \frac{\partial}{\partial \mu_{0a}} 
\left[
\Delta  \mu_{0a} D_a^p ({\bf z}_a) C_{ab}^p [f_a, f_b]({\bf z}_a)
\right]
\right]_{{\bf z}_a = {\bf Z}_a} \right\rangle_{\xi_a}
,
\nonumber \\ & & 
\end{eqnarray}
  where $\Delta v_{\parallel a}$ and  
$\Delta \mu_{0a}$ are written as 
\begin{eqnarray}
\label{DvDm}
& & 
\Delta  {\bf x}_a = 
- \mbox{\boldmath$\rho$}_a + \Delta  {\bf x}_a^{(2)}
,
\hspace*{5mm}
\Delta v_{\parallel a} =
\Delta v_{\parallel a}^{(1)}
+ \Delta v_{\parallel a}^{(2)}
, \nonumber \\
& & 
\Delta \mu_{0a} =
\Delta \mu_{0a}^{(1)}
+ \Delta \mu_{0a}^{(2)}
,
\end{eqnarray}
   and $f_a$ is given from $F_a$ by 
 $f_a({\bf z}_a) = 
F_a({\bf x}_a + \Delta  {\bf x}_a, 
v_{\parallel 0a} + \Delta v_{\parallel a},
\mu _{0a} + \Delta \mu_{0a})$.
   Here, 
$\Delta v_{\parallel a}^{(1)}$,  and $\Delta \mu_{0a}^{(1)}$ 
are the ${\cal O}(\delta)$ parts of 
$\Delta v_{\parallel a}$ and 
$\Delta \mu_{0a}$ given in Eq.~(\ref{Deltaz}). 
    In this Appendix, we do not derive 
expressions for the ${\cal O}(\delta^2)$ parts 
$\Delta  {\bf x}_a^{(2)}$, 
$\Delta v_{\parallel a}^{(2)}$, and $\Delta \mu_{0a}^{(2)}$ 
by the Lie perturbation expansion method which is used to define 
the gyrocenter coordinates 
with the well-conserved magnetic moment 
because it would unnecessarily give higher-order accuracy to the coordinate transformation 
than the accuracy of the gyrocenter motion equations themselves 
shown in Eqs.~(\ref{dXdt})--(\ref{dxi}). 
   Instead, we determine these ${\cal O}(\delta^2)$ terms from the conditions 
that 
the collisional change rates of energy and canonical toroidal angular momentum 
per unit volume in the gyrocenter space 
can be given in the conservative forms as shown below. 
   Thus, the ${\cal O}(\delta^2)$ terms are introduced not for accuracy of higher order 
in $\delta$ but for satisfying the conservation property of the collision operator. 

  In Eq.~(\ref{DCg2}),  the expansions in 
$(\Delta  {\bf x}_a^{(2)}, 
\Delta v_{\parallel a}, \Delta \mu_{0a})$ are truncated 
to the first order while the infinite series expansion in 
$\Delta {\bf x}_a^{(1)} \equiv - \mbox{\boldmath$\rho$}_a$ is retained 
because  fluctuations' wavelengths in the directions perpendicular to 
the equilibrium magnetic field can be of order of  the gyroradius $\rho_a$. 
   In the WKB (or ballooning) representation,  
the above-mentioned infinite series expansion can be treated 
using the Bessel functions of the gyroradius normalized by the perpendicular 
wavelength.~\cite{rutherford,AL,Sugama1996}  
   We should also note that the gyrophase average $\langle \cdots \rangle_{\xi_a}$ 
is taken so that the gyrokinetic equation with the collision term is solved only 
for the gyrophase-averaged part of the gyrocenter distribution function.   

   For an arbitrary function ${\cal A}_a^g ({\bf Z}_a)$ which is independent 
of the gyrophase $\xi_a$, we obtain the following formula,
\begin{eqnarray}
\label{DCgA2}
& & 
D_a^g ({\bf Z}_a)
\langle C_{ab}^g [F_a, F_b]({\bf Z}_a) \rangle_{\xi_a} {\cal A}_a^g ({\bf Z}_a)
\nonumber \\ 
& = & 
\left\langle 
\left[ 
D_a^p ({\bf z}_a) C_{ab}^p [f_a, f_b]({\bf z}_a) 
{\cal A}_a^p ({\bf z}_a)
+ 
\sum_{n=1}^\infty \frac{1}{n!}
\right. \right.
\nonumber \\ 
&  & \mbox{} \times
\sum_{i_1,\cdots,i_n}
\frac{\partial^n 
\left[
\rho_a^{i_1} \cdots \rho_a^{i_n}
D_a^p ({\bf z}_a)
C_{ab}^p [f_a, f_b]({\bf z}_a) {\cal A}_a^{p0} ({\bf z}_a)
\right]
}{
\partial x_a^{i_1} \cdots \partial x_a^{i_n}}
\nonumber \\ & & 
\mbox{}
- \frac{\partial}{\partial {\bf x}_a} \cdot
\left[
\Delta  {\bf x}_a^{(2)} D_a^p ({\bf z}_a) C_{ab}^p [f_a, f_b]({\bf z}_a)
{\cal A}_a^g ({\bf z}_a)
\right]
\nonumber \\ & & 
\mbox{}
- \frac{\partial}{\partial v_{\parallel a}} 
\left[
\Delta  v_{\parallel a} D_a^p ({\bf z}_a) C_{ab}^p [f_a, f_b]({\bf z}_a)
{\cal A}_a^g ({\bf z}_a)
\right]
\nonumber \\ & & 
\left. \left. 
\mbox{}
- \frac{\partial}{\partial \mu_{0a}} 
\left[
\Delta  \mu_{0a} D_a^p ({\bf z}_a) C_{ab}^p [f_a, f_b]({\bf z}_a)
{\cal A}_a^g ({\bf z}_a)
\right]
\right]_{{\bf z}_a = {\bf Z}_a} \right\rangle_{\xi_a}
, 
\nonumber \\ & & 
\end{eqnarray}
     where ${\cal A}_a^p ({\bf z}_a)$ and ${\cal A}_a^{p0} ({\bf z}_a)$ are 
defined by 
\begin{eqnarray}
\label{App}
{\cal A}_a^p ({\bf z}_a) & = & 
{\cal A}_a^{p0} ({\bf z}_a) 
+ 
\left( \Delta  {\bf x}_a^{(2)} \cdot \frac{\partial}{\partial {\bf x}_a}
+ \Delta  v_{\parallel a} \frac{\partial}{\partial v_{\parallel a}} 
\right.
\nonumber \\ & & 
\left. \mbox{}
+ \Delta  \mu_{0a}  \frac{\partial}{\partial \mu_{0a}} 
\right)
{\cal A}_a^g ({\bf z}_a)
,
\nonumber \\ 
{\cal A}_a^{p0} ({\bf z}_a) 
& = & 
{\cal A}_a^g ( {\bf x}_a - \mbox{\boldmath$\rho$}_a, 
v_{\parallel a}, \mu_{0a}) 
. 
\end{eqnarray}
We should note 
that the function ${\cal A}^p({\bf z}_a)$ defined from 
${\cal A}_a^g ({\bf Z}_a)$ 
in Eq.~(\ref{App}) does not exactly 
coincide with that given in Eq.~(\ref{Ap_series}) in Appendix~A by 
the second-  and higher-order terms in the series expansion with respect to 
$\Delta {\bf z}_a$. 
   Integrating Eq.~(\ref{DCgA2}) over the gyrocenter velocity space, 
we immediately obtain 
\begin{eqnarray}
\label{DCgA3}
& & 
\int dU \int d\mu  \int d\xi \; 
D_a^g ({\bf Z})
C_a^g ({\bf Z}) {\cal A}_a^g ({\bf Z})
\nonumber \\ 
& = & 
\left[ 
\int d^3 {\bf v} \; C_a^p ({\bf z}) 
{\cal A}_a^p ({\bf z})
\right]_{{\bf x} = {\bf X}} 
- \nabla \cdot {\bf J}_{Aa}^{\rm C}
, 
\nonumber \\ & & 
\end{eqnarray}
     where the transport flux ${\bf J}_{Aa}^{\rm C}$ due to collisions and finite gyroradii 
is defined by 
\begin{eqnarray}
\label{JAC3}
& & 
{\bf J}_{Aa}^{\rm C}
({\bf X})
\nonumber \\ 
&  = & 
\left[
\int d^3 {\bf v} \; 
[ - \mbox{\boldmath$\rho$}_a + \Delta {\bf x}_a^{(2)} ]
C_a^p ({\bf z})
{\cal A}_a^p ({\bf z})
\right]_{{\bf x} = {\bf X}}
\nonumber \\ 
&  & 
+
\sum_{n=1}^\infty 
\frac{-1}{(n+1)!}
\sum_{i_1,\cdots,i_n}
\frac{\partial^n}{
\partial X^{i_1} \cdots \partial X^{i_n}}
\nonumber \\ 
&  & 
\times
\left[
\int d^3 {\bf v} \; \mbox{\boldmath$\rho$}_a
\rho_a^{i_1} \cdots \rho_a^{i_n}
C_a^p ({\bf z})
{\cal A}_a^p ({\bf z})
\right]_{{\bf x} = {\bf X}}
.
\end{eqnarray}
   To the lowest order in $\delta$, 
   Eqs.~(\ref{DCgA2}), (\ref{DCgA3}), and (\ref{JAC3}) 
derived from the approximate collision operator in Eq.~(\ref{DCg2}) 
agree with 
Eqs.~(\ref{DCgA}), (\ref{intvDCgA}), and (\ref{JAC}) given 
in Appendix~B, respectively.  
   The particle flux $\mbox{\boldmath$\Gamma$}_a^{\rm C}$ due to 
collisions and finite gyroradii  is given from Eq.~(\ref{JAC3}) 
with putting ${\cal A}_a^p ({\bf z}) = 1$ 
in the same way as in Eqs.~(\ref{GammaC}).

   Now, let us take ${\cal A}_a^g ({\bf Z}) = H_a ({\bf Z})$ in Eq.~(\ref{DCgA3}). 
Here, $H_a ({\bf Z})$ denotes the gyrocenter Hamiltonian defined by 
Eq.~(\ref{Ha}). 
   It is desirable that the gyrocenter velocity-space integral 
$
\sum_a
\int dU \int d\mu \int d\xi \: 
D_a^g ({\bf Z})
C_a^g ({\bf Z}) 
H_a ({\bf Z})
$
takes the conservative form, which implies that the integral 
is expressed by the divergence term only and 
\begin{equation}
\label{CHp=0}
\sum_a \int d^3 {\bf v} \; C_a^p ({\bf z})
H_a^p ({\bf z}) =  0
\end{equation}
holds. 
       Here, using Eqs.~(\ref{Ha}) and (\ref{App}), 
$H_a^p ({\bf z})$ is given by 
\begin{equation}
\label{HZz}
H_a^p ({\bf z}_a) 
=
\frac{1}{2}m_a v_{\parallel a}^2 
+ \mu B_0 ({\bf x}_a) + e_a \phi ({\bf x}_a)
+ \Delta H_a ({\bf z}_a) 
,
\end{equation}
where 
\begin{eqnarray}
\label{DHz}
& & 
\hspace*{-8mm}
\Delta H_a ({\bf z}_a) 
 \equiv  
\left( 
\Delta {\bf x}_a^{(2)} \cdot \frac{\partial}{\partial {\bf x}_a} 
+
\Delta  v_{\parallel a}  \frac{\partial}{\partial v_{\parallel a}} 
+ \Delta  \mu_{0a}  \frac{\partial}{\partial \mu_{0a}} 
\right)
H_a ({\bf z}_a)
\nonumber \\ 
&  & 
\hspace*{15mm}
\mbox{}
+ \mu_{0a} 
\left[ 
B_0 ({\bf x}_a - \mbox{\boldmath$\rho$}_a) - B_0 ({\bf x}_a) 
\right]
\nonumber \\ 
&  & 
\hspace*{15mm}
\mbox{}
+ e_a 
\left[ 
\Psi ({\bf x}_a - \mbox{\boldmath$\rho$}_a, v_{\parallel a} , \mu_{0a})
- \phi ({\bf x}_a) 
\right]
.
\end{eqnarray}
   It is easily seen that Eq.~(\ref{CHp=0}) is satisfied if $\Delta H_a ({\bf z}_a) = 0$. 
   Then, substituting Eq.~(\ref{DvDm}) into Eq.~(\ref{DHz}) and 
using $\Delta H_a ({\bf z}_a) = 0$, we have 
\begin{eqnarray}
\label{aa1}
& &
\left(
\Delta {\bf x}_a^{(2)} \cdot \frac{\partial}{\partial {\bf x}_a} 
+
\Delta v_{\parallel a}^{(2)} \frac{\partial}{\partial v_{\parallel a}} 
+ \Delta \mu_{0a}^{(2)}  \frac{\partial}{\partial \mu_{0a}} 
\right)
H_a ({\bf z}_a)
\nonumber \\ 
& & 
\hspace*{-5mm}
=
- \mu_{0a} \left[ B_0 ({\bf x}_a - \mbox{\boldmath$\rho$}_a ) 
- B_0 ({\bf x}_a) + ( \mbox{\boldmath$\rho$}_a \cdot \nabla B_0 )
\left( 1 + \frac{e_a}{B_0}  
\right. \right.
\nonumber \\ 
& & \mbox{} 
\left. \left. \times \frac{\partial \Psi}{\partial \mu_{0a}} 
\right) \right]
- \frac{e_a^2}{B_0} \left[  \phi ({\bf x}_a) - 
\langle \psi ({\bf X}_a + \mbox{\boldmath $\rho$}_a) \rangle_{\xi_{a}}
\right] \frac{\partial \Psi}{\partial \mu_{0a}}  
\nonumber \\ & & \mbox{}
\hspace*{-5mm}
- \frac{e_a^2}{2m_a c^2}   \left\langle 
|{\bf A}_1 
({\bf X}_a + \mbox{\boldmath $\rho$}_a) 
|^2 \right\rangle_{\xi_a}
+ \frac{e_a^2}{2 B_0}
\frac{\partial}{\partial \mu} \langle 
[
\widetilde{\psi}_a ({\bf X}_a + \mbox{\boldmath $\rho$}_a)
]^2
\rangle_{\xi_a} 
\nonumber \\ 
& & \mbox{} 
+ e_a \left[ v_{\parallel a} {\bf b}\cdot\nabla {\bf b} \cdot 
\mbox{\boldmath$\rho$}_a 
+ \frac{1}{4}
( 3 \mbox{\boldmath$\rho$}_a \cdot \nabla {\bf b} \cdot
{\bf v}_\perp - {\bf v}_\perp \nabla {\bf b} \cdot 
\mbox{\boldmath$\rho$}_a ) 
\right.
\nonumber \\ & & 
\left. 
\mbox{}
- \frac{e_a}{m_a c} A_{1\parallel} \right]
\left( 
\frac{\partial}{\partial v_{\parallel a }}
- \frac{m_a v_{\parallel a}}{B_0} 
\frac{\partial}{\partial \mu_{0a}}
\right) \Psi
.
\end{eqnarray}
    We find that the right-hand side of Eq.~(\ref{aa1}) is of ${\cal O}(\delta^2)$.  
    Then, as remarked after Eq.~(\ref{aa3}), 
we can choose $\Delta {\bf x}_a^{(2)}$, 
$\Delta  v_{\parallel a}^{(2)}$, and 
$\Delta  \mu_{0a}^{(2)}$ which  
satisfy Eq.~(\ref{aa1}) and are of ${\cal O}(\delta^2)$ 
 so as to be consistent with Eq.~(\ref{Deltaz}). 

    When we use $\Delta {\bf x}^{(2)} =  0$, 
$\Delta v_{\parallel a} = \Delta v_{\parallel a}^{(1)}$, 
and $\Delta \mu_{0a} = \Delta \mu_{0a}^{(1)}$ for Eq.~(\ref{DCg2}) by putting 
$\Delta v_{\parallel a}^{(2)} =  \Delta \mu_{0a}^{(2)}= 0$, 
we have $\Delta H_a  ({\bf z}_a) = {\cal O}(\delta^2)$ and 
$\sum_a \int d^3 {\bf v} \; C_a^p ({\bf z}) H_a^p ({\bf z}) =  {\cal O}(\delta^3)$ 
because $C_a^p ({\bf z}) = {\cal O}(\delta)$ holds for the distribution function, the zeroth order 
of which is given by the local Maxwellian. 
    Therefore, even for this case where 
$
\sum_a
\int dU \int d\mu \int d\xi \;  
D_a^g ({\bf Z})
C_a^g ({\bf Z}) 
H_a ({\bf Z})
$ 
is not completely given in the conservative form, 
the residual term 
$\sum_a \int d^3 {\bf v} \; C_a^p ({\bf z}) H_a^p ({\bf z}) =  {\cal O}(\delta^3)$ 
is smaller by a factor of $\delta$ than other transport terms of ${\cal O}(\delta^2)$ 
in the lowest-order energy balance equation given by Eq.~(\ref{Econs3}) in Sec.~VI.B. 

    We next put 
${\cal A}_a^g ({\bf Z}) = (p_\zeta^c)_a^g ({\bf Z})$ in Eq.~(\ref{DCgA3}). 
Here, $(p_\zeta^c)_a^g ({\bf Z})$ denotes the canonical 
toroidal angular momentum defined by 
\begin{equation}
\label{pzcg}
(p_\zeta^c)_a^g({\bf Z})
\equiv
\frac{e_a}{c} A_\zeta^* ({\bf Z})
\equiv
\frac{e_a}{c} A_{0\zeta}({\bf X}) 
+ m_a U b_\zeta ({\bf X})
, 
\end{equation}
where $A_{0\zeta} = - \chi$ and $b_\zeta = I/B_0$. 
    We now see that 
$
\sum_a
\int dU \int d\mu \int d\xi \;  
D_a^g ({\bf Z})
C_a^g ({\bf Z}) 
(p_\zeta^c)_a^g({\bf Z})
$
  takes the conservative form if  
\begin{equation}
\label{Cpp=0}
\sum_a \int d^3 {\bf v} \; C_a^p ({\bf z})
(p_\zeta^c)_a^p ({\bf z}) =  0
. 
\end{equation}
Here, using Eqs.~(\ref{App}) and (\ref{pzcg}), $(p_\zeta^c)_a^p ({\bf z})$ is given by 
\begin{equation}
\label{pzcp}
(p_\zeta^c)_a^p ({\bf z}) 
=
\frac{e_a}{c} 
[A_{0\zeta}({\bf x}) + A_{1\zeta}({\bf x}) ]
+ m_a  v_\zeta 
+ \Delta (p_\zeta^c)_a ({\bf z}) 
,
\end{equation}
   where 
\begin{eqnarray}
\hspace*{-3mm}
\Delta (p_\zeta^c)_a ({\bf z}_a) 
& = & 
\left( 
\Delta {\bf x}_a^{(2)}  \frac{\partial}{\partial {\bf x}_a} 
+ 
\Delta  v_{\parallel a}  \frac{\partial}{\partial v_{\parallel a}} 
\right)
(p_\zeta^c)_a^g ( {\bf x}_a , v_{\parallel a})
\nonumber \\ 
&  & 
\mbox{} 
+ (p_\zeta^c)_a^g ( {\bf x}_a - \mbox{\boldmath$\rho$}_a, 
v_{\parallel a}) 
\nonumber \\ 
&  & 
\mbox{}
- \frac{e_a}{c} 
[A_{0\zeta}({\bf x}_a) + A_{1\zeta}({\bf x}_a) ]
- m_a  v_{\zeta a}
.
\end{eqnarray}
    Again, we easily see that Eq.~(\ref{Cpp=0}) is satisfied if 
$\Delta (p_\zeta^c)_a ({\bf z}_a) = 0$. 
    The ${\cal O}(\delta^2)$ variables, $\Delta  {\bf x}_a^{(2)}$ and 
$\Delta  v_{\parallel a}^{(2)}$, which 
meet the condition that  $\Delta (p_\zeta^c)_a ({\bf z}_a) = 0$,    
are given by 
\begin{eqnarray}
\label{aa2}
& &
\hspace*{-5mm}
\Delta  {\bf x}_a^{(2)} \cdot  \nabla
(p_\zeta^c)_a^g ({\bf z}_a) 
\nonumber \\
& = & 
-\frac{e_a}{c} ( \mbox{\boldmath$\rho$}_a  \cdot \nabla \chi )
\left[ 
\frac{v_{\parallel a}}{\Omega_a} {\bf b} \cdot (\nabla \times {\bf b} )
- \frac{1}{2 B_0} \mbox{\boldmath$\rho$}_a  \cdot \nabla B_0 
\right]
\nonumber \\ & & 
\mbox{} 
- \frac{m_a c}{e_a} \mu_{0a} W_\zeta
, 
\end{eqnarray}
   and 
\begin{eqnarray}
\label{aa3}
& &
\hspace*{-5mm}
\Delta  v_{\parallel a}^{(2)}
\frac{\partial}{\partial v_{\parallel a}} 
(p_\zeta^c)_a^g ({\bf z}_a) 
=
m_a b_\zeta ({\bf x}_a) \Delta  v_{\parallel a}^{(2)}
\nonumber \\ 
& = & 
-m_a v_{\parallel a} \left[ 
b_\zeta ({\bf x}_a - \mbox{\boldmath$\rho$}_a  ) 
 - b_\zeta ({\bf x}_a) 
 + \mbox{\boldmath$\rho$}_a \cdot \nabla b_\zeta  \right]
\nonumber \\ & & 
\hspace*{-5mm}
\mbox{} 
+ \frac{e_a}{c} \left[ \chi ({\bf x}_a - \mbox{\boldmath$\rho$}_a  ) 
- \chi ({\bf x}_a ) 
+ \mbox{\boldmath$\rho$}_a \cdot \nabla \chi
- \frac{1}{2} \mbox{\boldmath$\rho$}_a  \mbox{\boldmath$\rho$}_a 
: \nabla \nabla \chi \right]
, 
\nonumber \\ & & 
\end{eqnarray}
where $\nabla \equiv \partial/\partial {\bf x}_a$ and 
$W_\zeta \equiv - (\nabla R \cdot \nabla \chi )/(RB_0)
+ \frac{1}{2} b_\zeta {\bf b} \cdot ( \nabla \times {\bf b} )$. 
   As a solution to Eq.~(\ref{aa2}), we can assume  
$\Delta  {\bf x}_a^{(2)}$ to be given in the form 
$\Delta  {\bf x}_a^{(2)} =  \Delta x_{a \chi}^{(2)} \nabla \chi$. 
   We should note that 
$\nabla \chi \cdot \nabla (p_\zeta^c)_a^g = {\cal O}(\delta^{-1})$
and 
$\partial (p_\zeta^c)_a^g /\partial v_{\parallel a}
= m_a b_\zeta = {\cal O} (\delta^0)$ while the right-hand sides of 
Eqs.~(\ref{aa2}) and (\ref{aa3}) are 
of ${\cal O}(\delta)$ and ${\cal O}(\delta^2)$, respectively. 
   Therefore, Eqs.~(\ref{aa2}) and (\ref{aa3}) give 
$\Delta  {\bf x}_a^{(2)}$ and $\Delta  v_{\parallel a}^{(2)}$,  
which are both 
of ${\cal O}(\delta^2)$,  consistently with Eq.~(\ref{Deltaz}). 
   Then, these $\Delta  {\bf x}_a^{(2)}$ and $\Delta  v_{\parallel a}^{(2)}$ 
are substituted into Eq.~(\ref{aa1}) to determine 
$\Delta \mu_{0a}^{(2)}$ of ${\cal O}(\delta^2)$ as well. 

    Thus, the collision operator, which has the desired conservation properties 
as well as  
the accuracy required for correct description of collisional 
transport of the energy and the toroidal angular momentum,  
is given by Eq.~(\ref{DCg2}), in which $\Delta {\bf x}_a^{(2)}$
$\Delta v_{\parallel a}$, and $\Delta \mu_{0a}$ are defined by 
Eqs.~(\ref{DvDm}), (\ref{aa1}), (\ref{aa2}), and (\ref{aa3}). 
   Using this collision operator, 
putting ${\cal A}^p ({\bf z}) =  \frac{1}{2}m_a v_{\parallel}^2 
+ \mu_{0} B_0 ({\bf x}) + e_a \phi ({\bf x})$ and 
${\cal A}_a^p ({\bf z}) = (e_a/c) [ A_{0\zeta}({\bf x}) 
+ A_{1\zeta}({\bf x}) ] + m_a v_\zeta$ 
in  Eq.~(\ref{JAC3}) 
and taking their summation over species $a$ define 
the transport fluxes ${\bf Q}^{\rm C}$ and ${\bf J}_{p\zeta}^{\rm C}$ of  the energy and the canonical toroidal angular momentum, respectively, 
which appear in the energy and toroidal angular momentum balance equations in Secs.~V.A and B 
[see Eqs.~(\ref{Econs2a}), (\ref{Q}), (\ref{tamc1a}), and (\ref{PiC})]. 
    In the definition of ${\bf Q}^{\rm C}$ mentioned above, 
the contribution of the potential energy part $e_a \phi$ is written as $\phi \sum_a e_a \mbox{\boldmath$\Gamma$}_a^{\rm C}$ which is smaller than the contribution of 
the kinetic energy part by a factor of $\delta$ because the classical particle fluxes 
represented by the lowest-order part of $\mbox{\boldmath$\Gamma$}_a^{\rm C}$ 
are intrinsically ambipolar. 
   Therefore, the energy flux ${\bf Q}^{\rm C}$ defined here agrees 
with Eq.~(\ref{JKE}) to the lowest order in $\delta$. 
   Regarding the entropy production discussed in Appendix~B, 
the positive definiteness of the entropy production rate [corresponding to the first 
term on the left-hand side of Eq.~(\ref{intvDCglogF})] is only approximately 
shown by using the present model collision operator in 
Eq.~(\ref{DCgA3}) with ${\cal A}_a^g ({\bf Z})= -  [ \log F_a ({\bf Z})+1]$
because ${\cal A}_a^p ({\bf z})= -  [\log f_a ({\bf z})+1]$ is not rigorously 
derived from Eq.~(\ref{App}) 
without the infinite series expansion in $\Delta {\bf z}_a$ as in 
Eq.~(\ref{Ap_series}).

\section{Derivation of Eqs.~(\ref{PiC_av}) and (\ref{PiaC}) }

    In this Appendix, it is shown how to derive 
Eqs.~(\ref{PiC_av}) and (\ref{PiaC})  
by using the collision operator given in Appendix~C. 
    On the right-hand side of Eq.~(\ref{PiC}) where 
the radial flux $(\Pi^{\rm C})^s$ of the toroidal angular momentum 
due to collisions and finite gyroradii is defined, the two types of fluxes 
${\bf j}^{\rm C} = \sum_a e_a \mbox{\boldmath$\Gamma$}_a^{\rm C}$ and 
${\bf J}_{p\zeta}^{\rm C} = \sum_a {\bf J}_{p\zeta a}^{\rm C}$ are 
evaluated by taking the summation of Eq.~(\ref{JAC3}) over species $a$ 
 with putting ${\cal A}_a^p ({\bf z}) = e_a$ and 
${\cal A}_a^p ({\bf z}) = (p_\zeta^c)_a^p ({\bf z})$, 
respectively. 
Here, 
$(p_\zeta^c)_a^p ({\bf z}) = (e_a/c)  [A_{0\zeta}({\bf x}) + A_{1\zeta}({\bf x}) ]
   + m_a  v_\zeta$ is used for the collision operator which conserves 
the toroidal angular momentum as explained in Appendix~C. 
   Consequently, the ensemble average of $(\Pi^{\rm C})^s$ is expressed 
explicitly up to ${\cal O}(\delta^2)$ as 
\begin{eqnarray}
\label{PiCsens}
& & 
\langle (\Pi^{\rm C})^s \rangle_{\rm ens}
= 
\left\langle 
\left(
{\bf J}_{p\zeta}^{\rm C}
+ \frac{\chi}{c} {{\bf j}_L^{\rm C} }
\right)
\cdot \nabla s
\right\rangle_{\rm ens}
\nonumber \\ 
&  = & 
\sum_a 
\int d^3 {\bf v} 
\langle C_a^p ({\bf z})  \rangle_{\rm ens}
\left[ 
\frac{e_a}{2 c}
\mbox{\boldmath$\rho$}_a
\mbox{\boldmath$\rho$}_a 
: \nabla \chi  \nabla s 
\right. 
\nonumber \\ 
&  & 
\left. 
\mbox{}
- 
m_a 
\mbox{\boldmath$\rho$}_a
{\bf v}_\perp 
: {\bf e}_\zeta  \nabla s  
\right]
+ {\cal O}(\delta^3)  
\nonumber \\ 
& =  & 
-
\sum_{a, b} \frac{m_a c |\nabla s|^2}{2 e_a B_0}
\frac{\partial \chi}{\partial s}
\int dU \int d \mu \int d\xi \;  D_a \mu
\nonumber \\ 
&   & 
\mbox{} \times
\left[
C_{ab}^p ( \langle F_{a1} \rangle_{\rm ens} , F_{bM} )
+ C_{ab}^p ( F_{aM}, \langle F_{b1} \rangle_{\rm ens}  )
\right]
\nonumber \\ 
&   & 
\mbox{}
+ {\cal O}(\delta^3) 
,
\end{eqnarray}
from which Eqs.~(\ref{PiC_av}) and (\ref{PiaC}) are immediately obtained. 
    It is noted that 
the ${\cal O}(\delta^2)$ part of $\langle (\Pi^{\rm C})^s \rangle_{\rm ens}$ 
has no contribution from the gyrophase-dependent part of the distribution 
function, the lowest-order part of which is given by 
$\widetilde{f}_{a1} = - \mbox{\boldmath$\rho$}_a \cdot \nabla F_{aM}$ 
with the gradient operator $\nabla$ taken for the fixed energy variable 
$\varepsilon = \frac{1}{2} m_a v^2 + e \langle \phi \rangle_{\rm ens}$.



\end{document}